\documentclass[aps,twocolumn,secnumarabic,nobalancelastpage,amsmath,amssymb,
nofootinbib]{revtex4-1}
\usepackage{dcolumn}
\usepackage{bm}
\usepackage[latin1]{inputenc}
\usepackage[active]{srcltx}%habilita a pesquisa inversa
\usepackage{ae,aecompl}
\usepackage{caption}
\usepackage{subcaption}
\usepackage{graphicx}
\usepackage{amsmath}
\usepackage{color}

\begin{document}

%\begin{frontmatter}
\title{Braiding of edge states in narrow zigzag graphene nanoribbons: effect of the third neighbors hopping}

%\author[Jorge]{J. H. Correa}
%\author[Figueira]{M. S. Figueira}
%\address[Jorge,Figueira]{Instituto de F\'{i}sica, Universidade Federal Fluminense, 24210-340, Niter\'oi, RJ, Brasil}

\author{J. H. Correa$^{1}$}
\email{jorgehuamani90@gmail.com}
\author{A. Pezo$^2$}
\author{M. S. Figueira$^1$}

\affiliation{$^{1}$Instituto de F\'{i}sica, Universidade Federal Fluminense, Av. Litor\^anea s/N, CEP: 24210-340, Niter\'oi, RJ, Brazil\\
$^2$Centro de Ci\^encias Naturais e Humanas, Universidade Federal do ABC, 
CEP: 09210-170, Santo Andr\'e, S\~ao Paulo, Brazil}

\begin{abstract}

 We study narrow zigzag graphene nanoribbons (ZGNRs), employing density functional theory (DFT) simulations and the tight-binding (TB) method. The main result of these calculations  is the  braiding of the conduction and valence bands, generating Dirac cones for non-commensurate wave vectors $\vec{k}$. Employing a TB Hamiltonian, we show that the  braiding is generated by the  third-neighbor hopping (N3). We calculate the band structure, the density of states and the conductance, new conductance channels are opened, and the conductance at the Fermi energy assumes integer multiples of the quantum conductance unit $G_{o} = 2e^{2}/h$. We also investigate the satisfaction of the Stoner criterion by these ZGNRs. We calculate the magnetic properties of the fundamental state employing LSDA (spin-unrestricted DFT) and we confirm that  ZGNRs with $N=(2,3)$ do not satisfy the Stoner criterion and as such the magnetic order could not be developed at their edges. These results are confirmed by both tight-binding and LSDA calculations.
\end{abstract}


\maketitle
%\end{frontmatter}

\section{Introduction}
\label{sec1}

The graphene era began with the seminal work of Andre K. Geim and Konstantin S. Novoselov \cite{Novoselov666}, who isolated sheets of graphite crystal only one atom thick.  At low temperatures, the graphene density of states exhibits a V-shaped gap, and at low energies its dispersion relation is linear. The electrons behave like massless fermionic particles, obeying the Dirac equation\cite{CastroNeto09}. However, graphene is gapless and cannot be used in microelectronics. Therefore, it is necessary to open and control  the  gap  without drastically changing its mobility. One effective way to  open a band gap is by employing electronic confinement, which is naturally present in the geometric structure of nanoribbons, turning these systems into excellent candidates to substitute silicon in technological applications\cite{Gunlycke207}. However, from the experimental point of view, their  synthesis produces nanoribbons with roughness at their edges that presents an adverse effect on their electronic transport properties\cite{Gunlycke07,Celis2016}. But this situation can change due to  a recent botton-up synthesis of $6$-ZGNR \cite{Pascal16}, (ZGNR with six carbon zigzag lines wide) with atomically precise zigzag edges.

Graphene zigzag nanoribbons (N-ZGNRs) (where $N$ represents the number of carbon zigzag lines along the width) exhibit several fundamental states. They can be metallic, insulating, or semiconducting, constituting a frontier in the research of graphene-based materials. Considering a nearest-neighbor hopping tight-binding calculation \cite{Nakada96}, the edge states of ZGNRs were theoretically predicted to couple ferromagnetically along the edge and antiferromagnetically between them, but direct observation of spin-polarized edge states of ZGNRs have not yet been achieved, owning to the limited precision of current measurement techniques \cite{Pascal16}.  

We develop a tight-binding calculation of ZGNRs taking into account only first- (NN) and third-neighbor (N3) hopping \cite{Kivelson1983,Kivelson2013,Korytar17}. Graphene ZGNRs exhibit localized edge states originating from the sublattice or chiral symmetry \cite{Ostaay11}, which is associated with the existence of two non-equivalent sublattices, $A$ and $B$. The lattice becomes bipartite, and a chiral symmetry is connected to the particle-hole symmetry of the band structure. When the N3 hopping is included in the calculations the Stoner criterion \cite{Ziman72,Teodorescu008} is not satisfied by low-width ZGNRs with $N=(2,3)$ and, therefore, these nanoribbons could not develop magnetic order at their edges. 

We do not consider second-neighbor hopping (N2),  between sites sharing the same sublattice, because its main effect is to break the particle-hole symmetry between the valence and conduction band \cite{Cristina2011} without introducing any qualitative changes in the braiding of the edge states as shown by our DFT calculations of section \ref{sec3}. However, N2 hopping becomes important when the spin-orbit interaction is taken into account in doped graphene with hollow adatoms \cite{Brey15}. A more complete study of the effects of the N2 hopping in the properties of the nanoribbons was reported in the Ref. \cite{Caio14}. 

The general effects of the N3 hopping in pristine graphene, nanoribbons and in the context of the Kane-Mele \cite{Kane_Mele_05} generalized model have been considered in several articles \cite{Reich2002,Gunlycke08,Hancock10,Wu2010,Kundu11,Cristina2011,Tran17,Hung13,Hung14,Chen15}, but their particular effects on the properties of zigzag nanoribbons with low $N$ width have not been addressed until now. The main objective of this paper is to fill up this gap employing the DFT and TB methods to calculate the band structure, the density of states, and the conductance of these kind of ZGNRs. To investigate the magnetic nature of their fundamental state we employ a spin-unrestricted LSDA calculation. On general grounds, both DFT and TB results are similar, showing that the inclusion of N3 hopping in the TB calculations is necessary to capture the physics of those low-width ZGNRs.

One important consequence of the inclusion of N3 hopping in the calculation is the lifting of the degeneracy of the edge states at the Fermi energy, producing braiding of the conduction and valence bands and generating Dirac cones for non-commensurate values of the wave vector $\vec{k}$ \cite{Kivelson2013,Korytar17}. New conductance channels are opened at the Fermi energy, with the conductance assuming integer multiples of the quantum conductance $G_{o} = 2e^{2}/h$ unit. The edges of the ZGNRs behave like a quantum wire, with the number of conductance channels being controlled by the width of the ZGNRs. It is also expected that the inclusion of N3 coupling qualitatively changes the properties of the edge states in pristine graphene \cite{Cristina2011}, carbon nanotubes \cite{Reich2002}, and armchair nanoribbons that include defects \cite{Wu2010}.

This paper is organized as follows: In sec. \ref{sec2}, we introduce the tight-binding Hamiltonian of the system that considers only NN and N3 hopping, and we discuss the Landauer-Buttiker formalism in order to calculate the conductance. In sec. \ref{sec3} we present the DFT results in the absence of spin polarization and obtain the braiding of the conduction and valence bands. In sec. \ref{sec4}, we introduce the tight-binding method and compare its results  with the corresponding DFT one, obtained in sec. \ref{sec3}. Employing TB we also calculate the transport properties and discuss their physical consequences. In sec.  \ref{sec5}, employing the spin polarized DFT (LSDA),  we discuss the magnetic order of the fundamental state and the fulfillment of the Stoner criterion. Finally,  in sec. \ref{sec6} we summarize the main findings of the paper and present the concluding remarks. 

\section{The model}
\label{sec2}

\begin{figure}[tbh]
\begin{center}
\includegraphics[clip,width=0.4\textwidth,angle=0.0]
{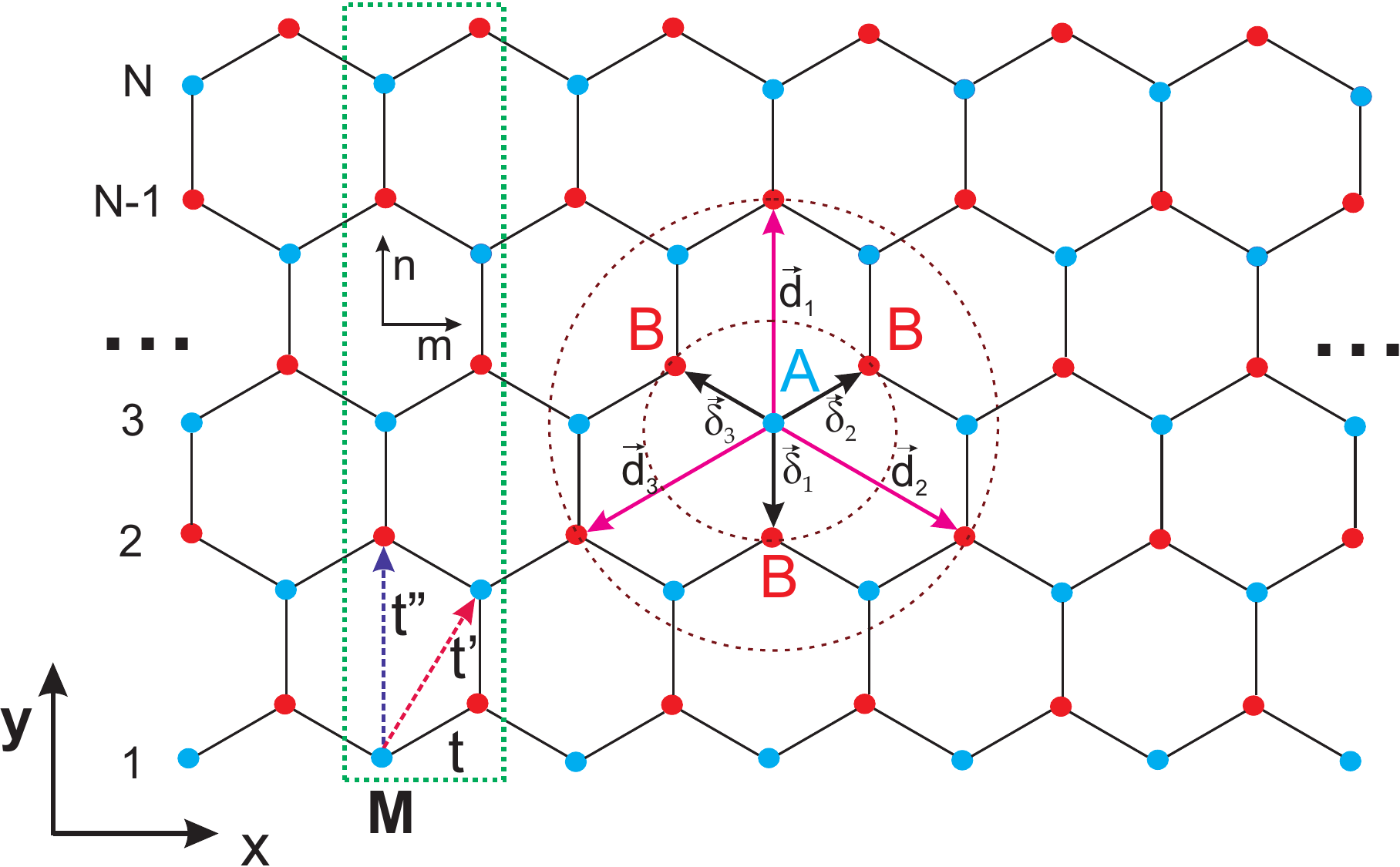}
\end{center}
\caption{(Color online) Schematic diagram of a generic ZGNR, where the green dashed region $M$ denotes the unit cell containing NN $(t)$, 
N2 $(t')$ and N3 $(t'')$ hopping.}
\label{zigzag_unitcell}
\end{figure}

The honeycomb lattice can be modeled by a tight-binding Hamiltonian \cite{Cristina2011}

$$
H =  -t\sum_{i,j=NN }c^{\dagger}_{iA}c_{jB} 
-t'\sum_{i,j=N2}(c^{\dagger}_{iA}c_{jA}+ c^{\dagger}_{iB}c_{jB}) 
$$
\begin{equation}
-t''\sum_{i,j=N3} c^{\dagger}_{iA}c_{jB} + H. c. ,
\label{H1}
\end{equation}
where $t$, $t'$, and $t''$ represent the NN, N2, and N3 hopping, respectively, $(i,j)$ labels the different sites of the unit cell, and $c^{\dagger}_{ij,AB}/c_{ij,AB}$ creates/annihilates an electron at site $(i,j)$ of the sub-lattice $(A,B)$ also, the negative sign of the hopping is associated with the formation of the proper bonding and anti-bonding state alignment of the $p_{z}$ orbitals in graphene. \cite{Yazyev2010}. In Fig. \ref{zigzag_unitcell}, we describe the different hopping between neighboring sites of the unit cell $M$ and represent the corresponding vectors of the first 
($\vec{\delta_{i}}$) and the third nearest neighbors ($\vec{d_{i}}$) as 
\begin{equation}
\vec{\delta}_{1} = \left(0,-a\right),\hspace{0.3cm}\vec{\delta}_{2} = \left(\frac{\sqrt{3}a}{2},\frac{a}{2}\right),\hspace{0.3cm}\vec{\delta}_{3} = \left(-\frac{\sqrt{3}a}{2},\frac{a}{2}\right) ,
\label{First}
\end{equation}
\begin{equation}
\vec{d}_{1} = \left(0,2a\right),\hspace{0.3cm}\vec{d}_{2} = \left(\sqrt{3}a,-a\right),\hspace{0.3cm}\vec{d}_{3} = \left(-\sqrt{3}a,-a\right) ,
\label{Third}
\end{equation}
where $a = 2.46$\AA \hspace{0.05cm} is the graphene lattice parameter. In order to obtain the ZGNR Hamiltonian that we use in our calculations, we follow reference \cite{Marcos2015}, and we write 
$$
H= -t\sum_{m} \sum_{n=1}^{N} \Bigl\{\vert A,m,n\rangle\langle B,m-1/2,n\vert+\nonumber \\
$$
$$
\vert A,m,n\rangle\langle B,m+1/2,n\vert\ + \vert A,m,n\rangle\langle B,m,n-1\vert\Bigl\}  \nonumber \\
$$
\begin{equation}
-t'' \sum_{m} \sum_{n=1}^{N}\vert A,m,n\rangle\langle B,m,n+1\vert + H.c. ,
\label{H2}
\end{equation}
where $A$ and $B$ indicate the non-equivalent sublattice sites of Fig. \ref{zigzag_unitcell}. The label $m$ runs over the unit cell along the infinite direction $x$, whereas the label $n=1,2, \cdots N$ is associated with the width $N$ of the nanoribbon along the $y$ direction. As discussed in the introduction, we do not consider the effect of N2 hopping, because it occurs between sites of the same sub-lattice and breaks the chiral symmetry of the lattice. Besides this, in order to simplify the calculations, we only consider N3 hopping along the $y$ direction, corresponding to the vector $\vec{d_{1}}$, and we discard N3 lateral hopping along $\vec{d_{2}}$ and $\vec{d_{3}}$, since it does not change the results in a qualitative way.
 
To derive the dispersion relations, we apply a Fourier transform to the ZGNR Hamiltonian, along the translationally invariant $x$-axis
$$
\vert\Psi\rangle = \frac{1}{\sqrt{M}}\sum_{m} \sum_{n=1}^{N} e^{i\vec{k}_{x}.\vec{R}_{m}}[ \psi_{A}(\vec{k},n)\vert A,m,n\rangle +\nonumber \\
$$
\begin{equation} 
\psi_{B}(\vec{k},n)\vert B,m,n\rangle ] ,
\label{function}
 \end{equation}
where $M$ labels the unit cell, $\vec{R}_{N}$ is the vector position of the site $N$, and $\vec{k}_{x}$ is the momentum along the $x$ axis. Substituting the Hamiltonian (Eq. \ref{H2}) and the eigenvectors equation (Eq. \ref{function}) into the Schr\"{o}dinger equation
\begin{equation}
H\vert\Psi\rangle = E\vert\Psi\rangle ,
\end{equation}
we obtain
$$
E\psi_{A}(\vec{k},n) = -t\left[2\psi_{B}(\vec{k},n)\cos\left( \frac{k_{x}a}{2}\right) + \psi_{B}(\vec{k},n-1)\right] \nonumber \\
$$
\begin{equation}
-t''\left[\psi_{B}(\vec{k},n+1)\right] ,
\label{Disp1}
\end{equation}

$$
E\psi_{B}(\vec{k},n) = -t\left[2\psi_{A}(\vec{k},n)\cos\left( \frac{k_{x}a}{2}\right) + \psi_{A}(\vec{k},n+1)\right] \nonumber \\
$$
\begin{equation}
 -t''\left[\psi_{A}(\vec{k},n+1)\right] .
\label{Disp2}
\end{equation}

The dispersion relations are obtained as the numerical solution of  Eqs. \ref{Disp1} and \ref{Disp2}.

\begin{figure}
    \centering
    \begin{subfigure}[t]{0.40\textwidth}
        \centering
        \includegraphics[width=\linewidth]{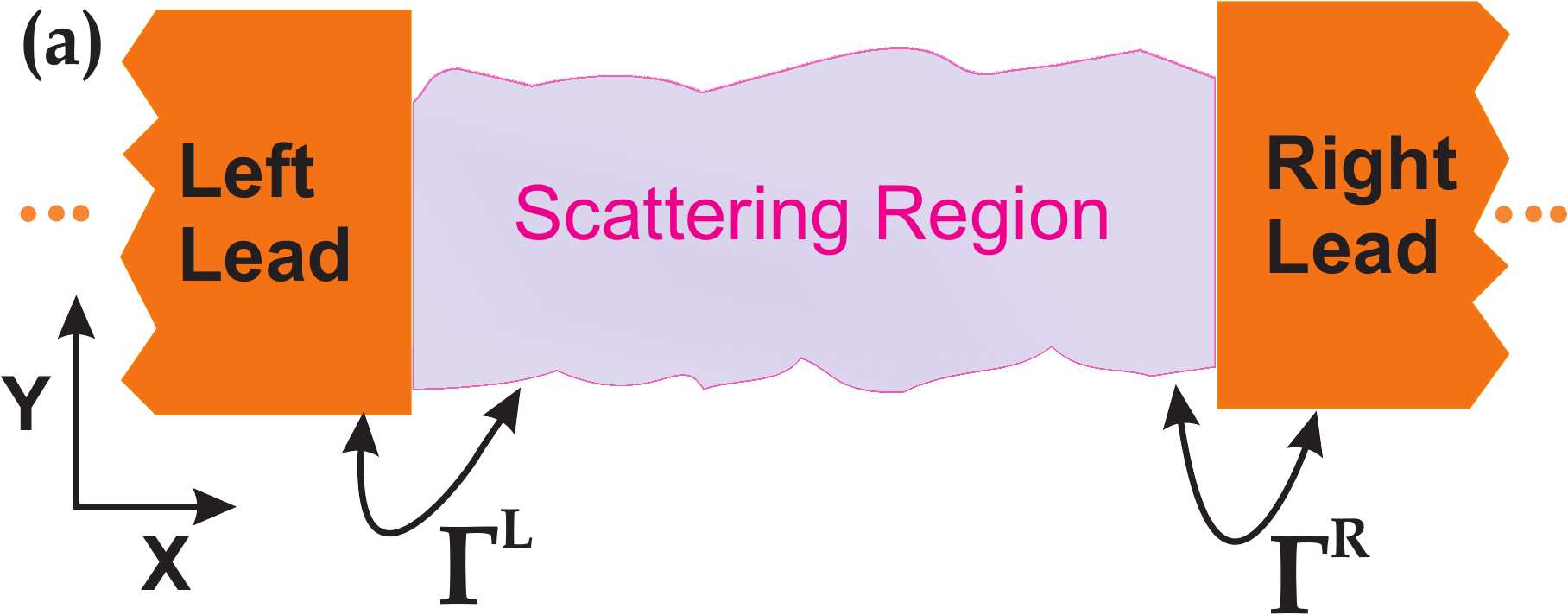} 
        %\caption{} 
			\label{zigzag_conductance}
    \end{subfigure}
    \begin{subfigure}[t]{0.40\textwidth}
    \centering
        \includegraphics[width=\linewidth]{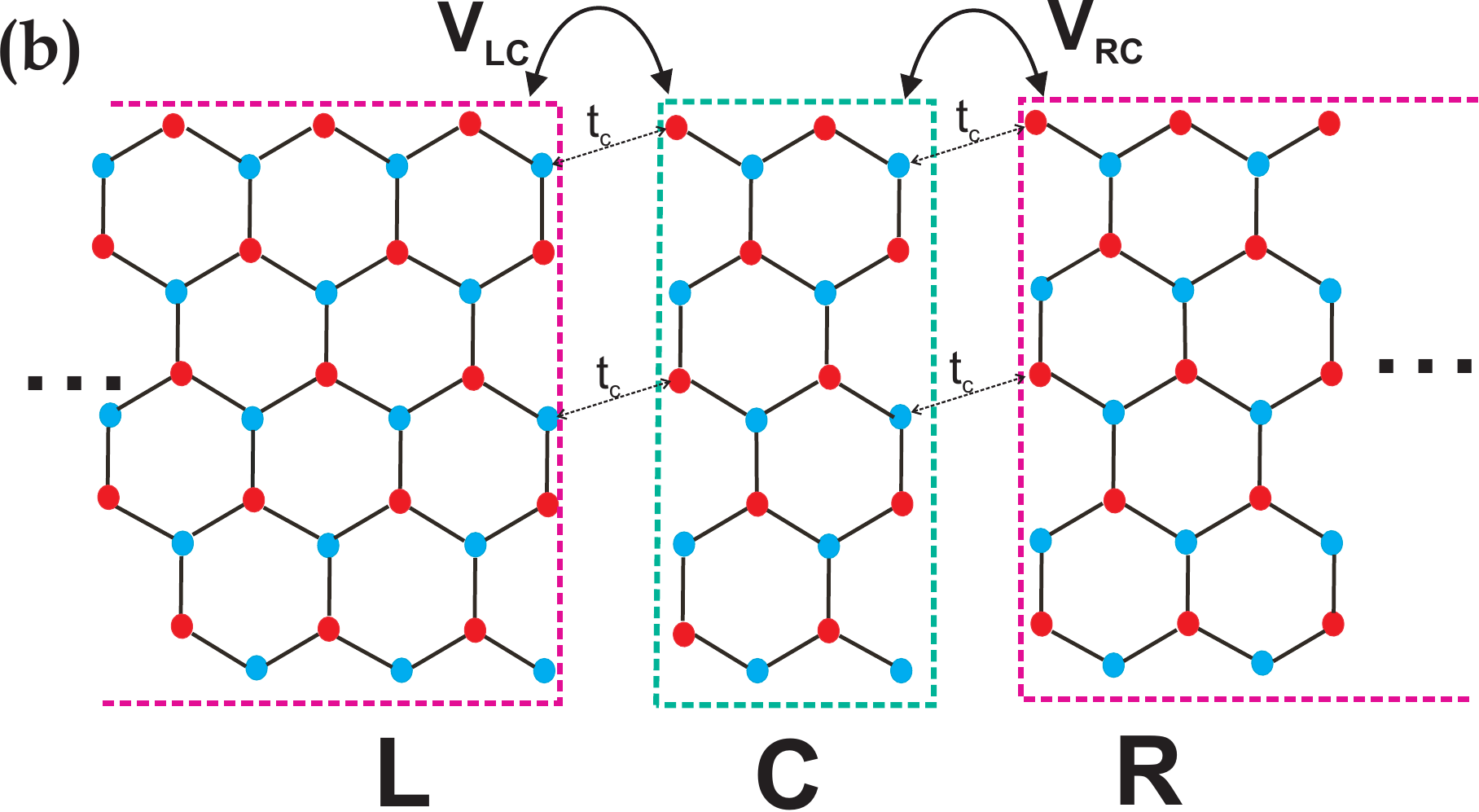} 
         %\caption{} 
				\label{duas_celulas}
						\end{subfigure}	
 \caption{(Color online) a) Schematic diagram of the two terminal  device, (scattering region) attached to equal semi-infinite leads through the coupling ($\Gamma_{L}, \Gamma_{R}$) functions. b) $V_{(L,R),C}$ is the matrix that couples the central part to the leads through the hopping 
$t_{c}$.} 
				\label{diagrama_condutancia}														
\end{figure}
In many systems, the non-equilibrium surface Green's function method \cite{Rubio}  is employed to calculate transport properties. The main advantages of this approach are its simplicity and its low computational cost. To apply this technique to the ZGNRs problem, the system is divided into three regions: the central or scattering region and the left and right leads, as shown in Fig. \ref{diagrama_condutancia}a. The influence of the leads on the central part is considered through their self-energies $(\Sigma_{L,R})$. The electrical conductance is calculated by employing the cell indicated in the rectangular region $C$ of Fig. \ref{diagrama_condutancia}b, through the Landauer-Buttiker formula \cite{ginetom,nardelli} 

\begin{equation}
 G(z) = G_{o}Tr\{\Gamma^{L}(z) {G}_{c}^{r}(z) \Gamma^{R}(z) {G}_{c}^{a}(z)\},  
\end{equation}
where $G_{o} = 2e^{2}/h$ is the quantum conductance unit. $Tr\{\cdots\}$ indicates the trace of the product of the retarded and advanced Green's functions of the central part ${G}_{c}^{r,a}(z)$, respectively, and its couplings to the leads $\Gamma^{j}$(z) $(j= L,R)$; $z=E \pm i\eta$, where $E$ is the energy and $\eta \rightarrow 0^{+}$ is an infinitesimal real quantity. To calculate ${G}_{c}^{r}(z)$ and ${G}_{c}^{a}(z)$, we employ $z=E - i\eta$ and $z=E + i\eta$, respectively. The Green's functions of the two-terminal device are given by
\begin{equation}
{G}_{c}^{r,a}(z) = [z - H_{c} - \Sigma_{L}(z) - \Sigma_{R}(z)]^{-1} ,
\label{green_conductor}
\end{equation}
where $H_{c}$ is the Hamiltonian of the central part, $\Sigma_{j}(z) =V^{\dagger}_{j,c}g_{j}(z)V_{j,c}$ $(j= L,R)$ are the self-energies, $V_{j,c}$ is the matrix that couples the central part to the leads through the hopping matrix $t_{c}$, and $g_{j}$ is the Green's function of the semi-infinite leads, which is calculated iteratively \cite{ginetom,nardelli,Rubio}. We define the couplings to the leads $\Gamma^{L,R}$ as
\begin{equation}
\Gamma^{j}(z) = i \left\lbrace \Sigma^{j}(z) - \left[ \Sigma^{j}(z)\right]^{\dagger}\right\rbrace \hspace{0.2cm} (j= L,R) .
\end{equation}
The local (LDOS) and the total density of states $\rho(E)$ are given  by
\begin{equation}
LDOS(E)=-(1/\pi){Im[{G}_{c}^{r}(z)_{ii}]},
\label{LDOS}
\end{equation}
\begin{equation}
\rho(E) = \frac{1}{L}\sum_{i=1}^{L} LDOS(E) ,
\label{totalLDOS}
\end{equation}
where $L=2N$ is the number of sites along the transverse direction.

\section{Density functional theory simulations}
\label{sec3}

In this section, we develop an $ab$ $initio$ calculation by means of the density functional theory (DFT) to investigate the  behavior of the band structure of narrow width zigzag nanoribbons. We employed in all the calculations, the local (spin) self-consistent pseudopotential method using the local density approximation (LDA) of the density functional theory (DFT) and, for the purposes of magnetism computations discussed in Sec. \ref{sec5}, the L(Spin)DA as implemented in the SIESTA package \cite{SIESTA}. An energy cutoff of $350$ Rydbergs (Ry) was employed with a double-($\zeta$) polarized basis set. In this approach, the electron density is obtained by integrating the density matrix with the proper Fermi-Dirac distribution. The geometry of each ZGNR was fully relaxed until the force felt by each atom was less than $0.001$ eV/A. A $k$-sampling of $20$ $k$-points was employed. We set the vacuum between edges and planes as $25$ and $20$ \AA \hspace{0.05cm}, respectively, in order to avoid spurious interactions with the periodic images. The edges of the ZGNRs were passivated with hydrogen to neglect effects related to carbon dangling bonds.

\begin{figure}[tbh]
\begin{center}
\includegraphics[clip,width=0.45\textwidth,angle=0.0]
{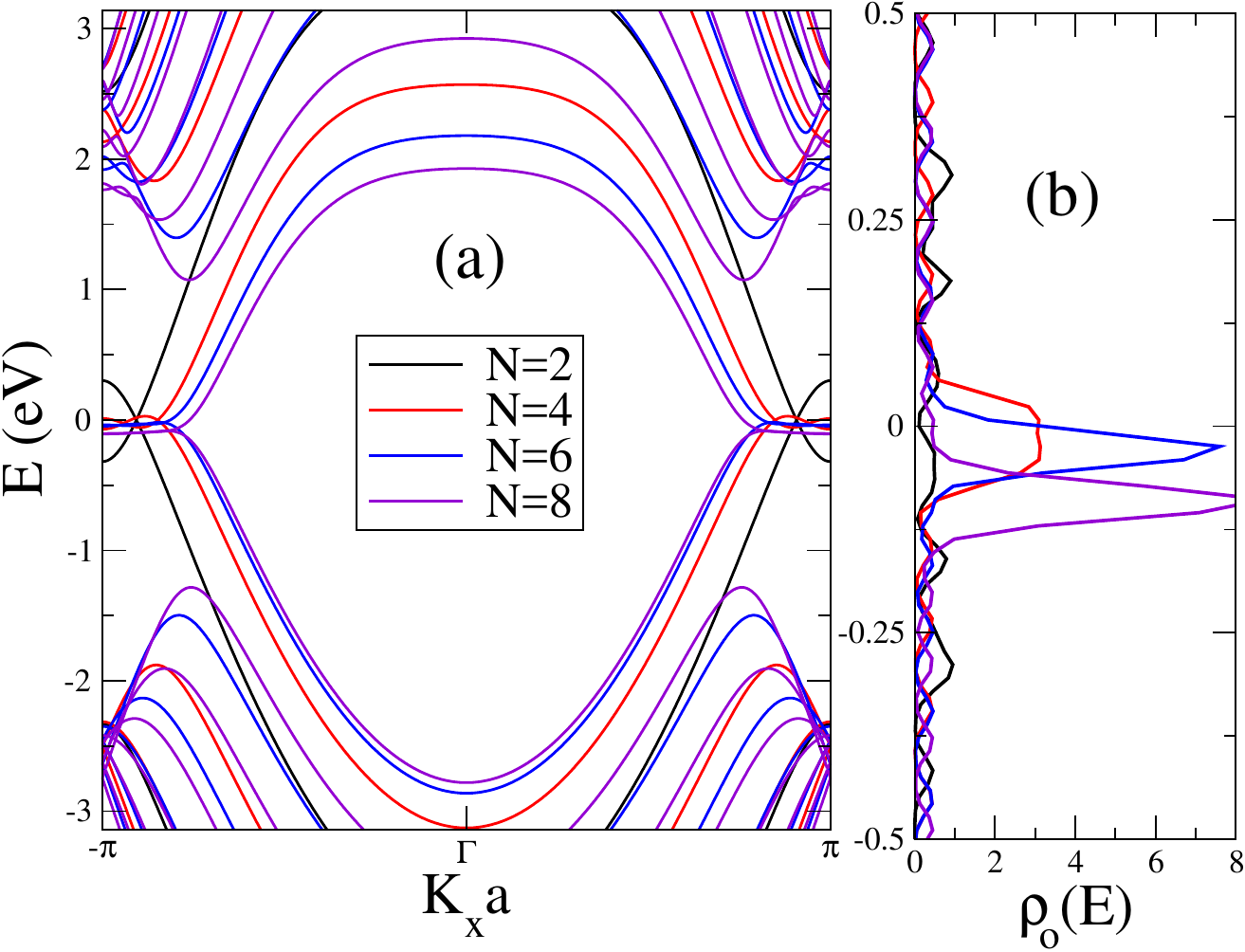}
\end{center}
\caption{(Color online) Electronic band structure (a) and unpolarized density of states (b) for ZGNRs with even width, $N=2,4,6,8$.}
\label{DFT_band_par}
\end{figure}

\begin{figure}[tbh]
\begin{center}
\includegraphics[clip,width=0.45\textwidth,angle=0.0]
{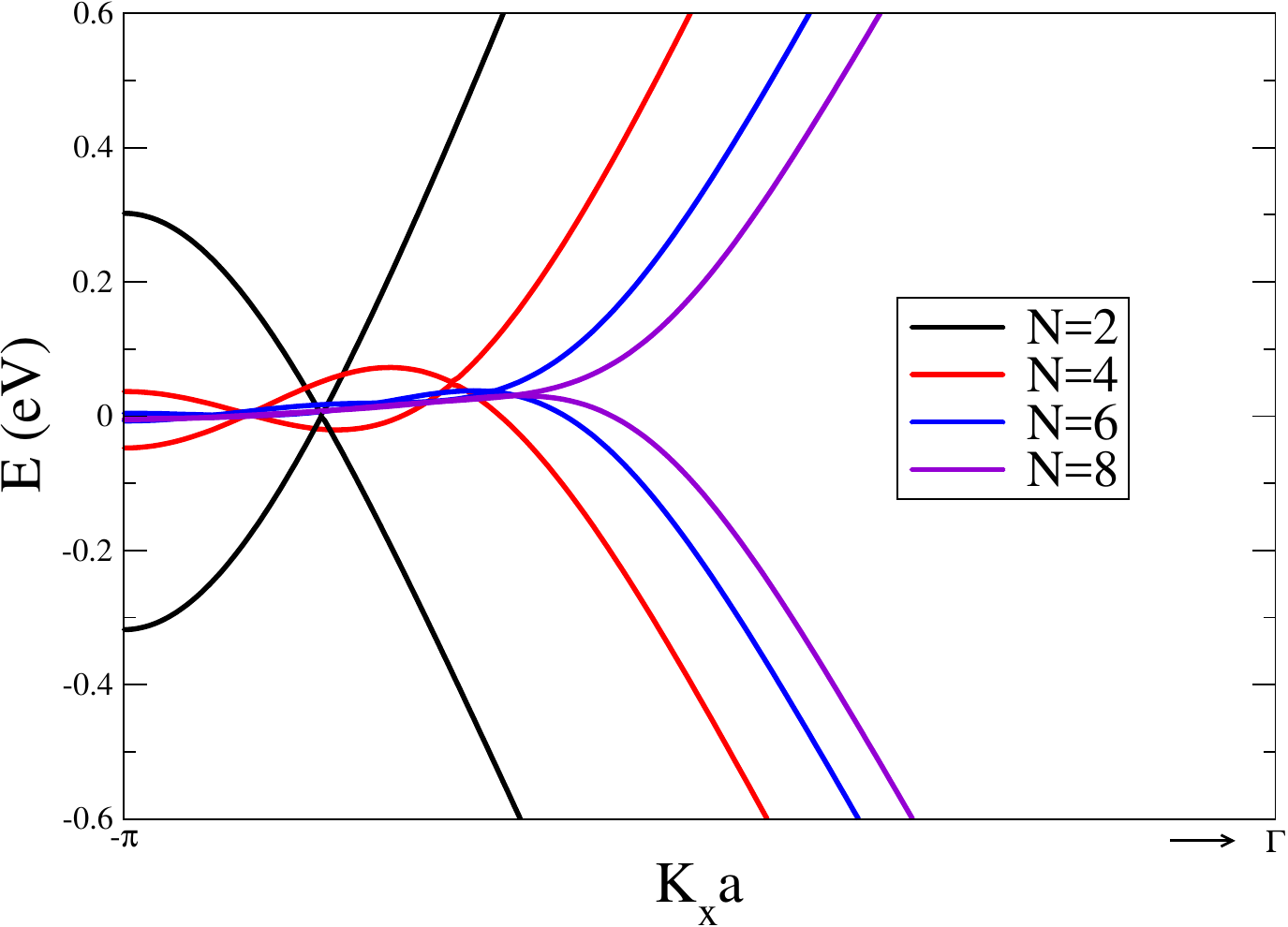}
\end{center}
\caption{(Color online) Zoom showing the existence of braiding of conduction and valence bands at the Fermi level, for ZGNRs with even width $N=2,4,6,8$.}
\label{Zoom_par}
\end{figure}

\begin{figure}[tbh]
\begin{center}
\includegraphics[clip,width=0.45\textwidth,angle=0.0]
{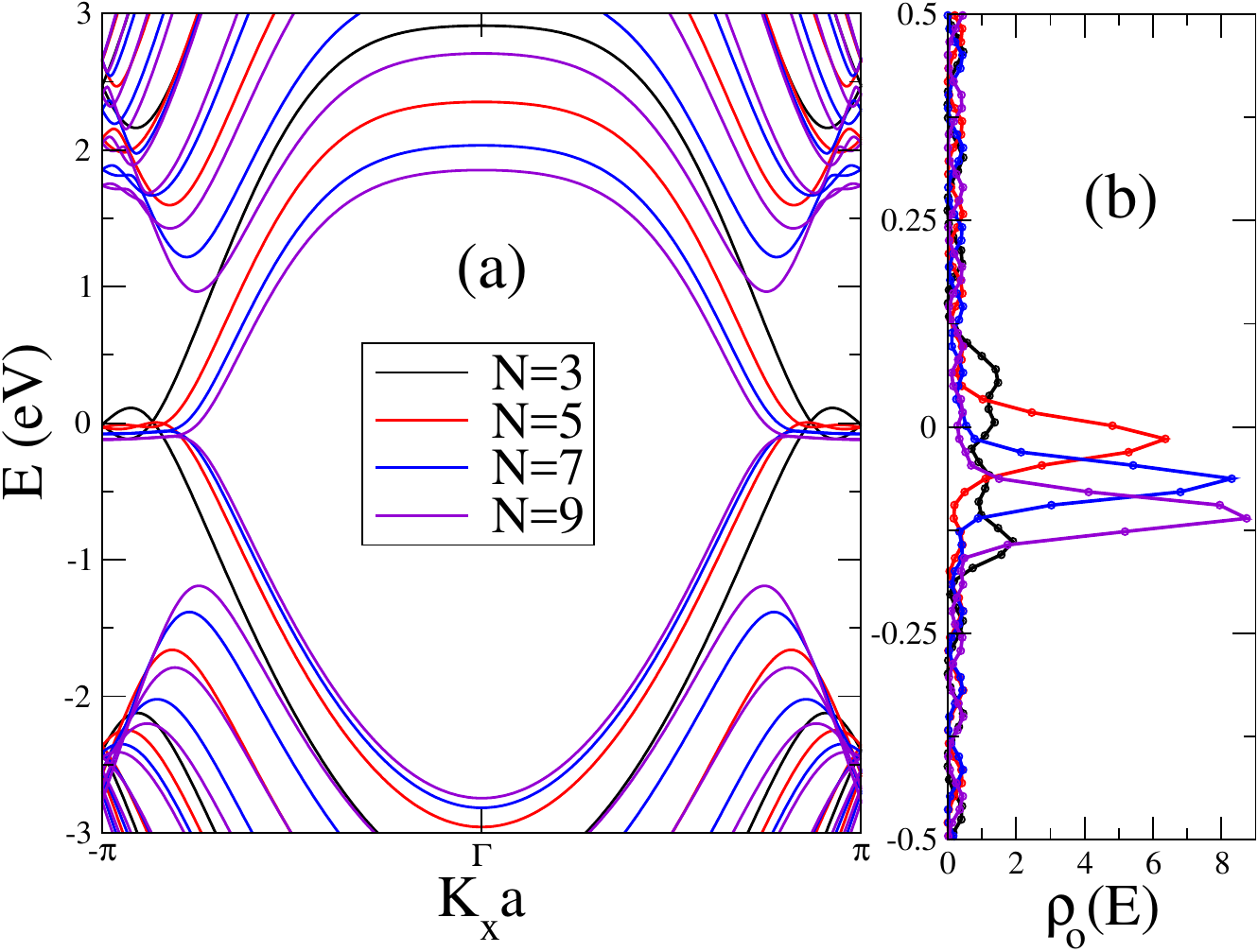}
\end{center}
\caption{(Color online) Electronic band structure (a) and unpolarized density of states (b) for ZGNRs with odd width, $N=3,5,7,9$.}
\label{DFT_band_impar}
\end{figure}
\begin{figure}
    \centering
    \begin{subfigure}[tbh]{0.40\textwidth}
        \centering
        \includegraphics[width=\linewidth]{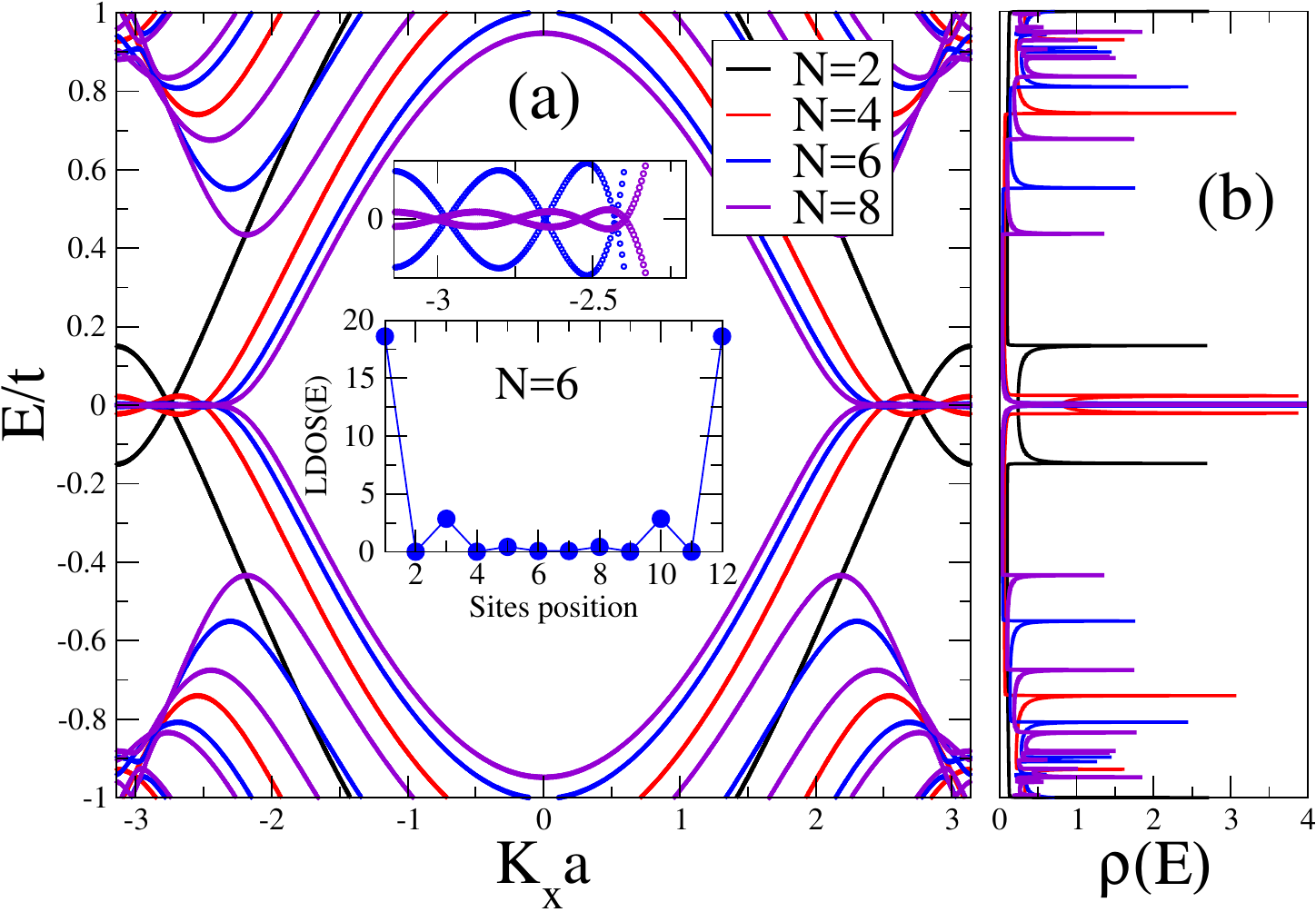} 
 			\label{band_dos}
    \end{subfigure}
		
			\vspace{0.5cm}
		\begin{subfigure}[tbh]{0.40\textwidth}
    \centering
        \includegraphics[width=\linewidth]{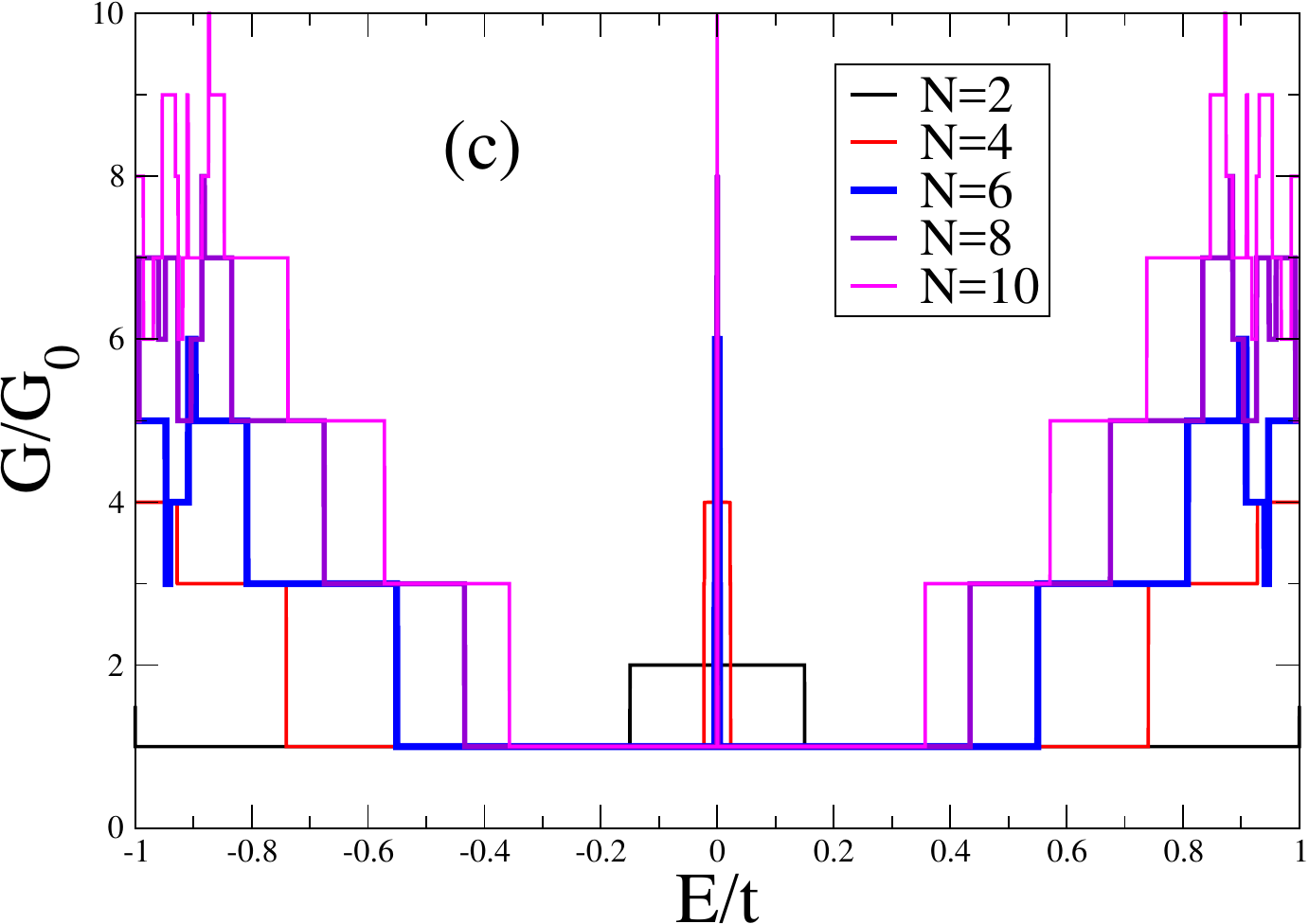} 
				\label{condutance_par}
						\end{subfigure}
				\caption{(Color online) a) Band structure,  b) density of states and c) conductance of ZGNRs with $t$ of the same sign as $t''$, and $N=2,4,6,8,10$.}
					\label{band_dos_cond_1}											
\end{figure}
\begin{figure}
    \begin{subfigure}[tbh]{0.40\textwidth}
        \includegraphics[width=\linewidth]{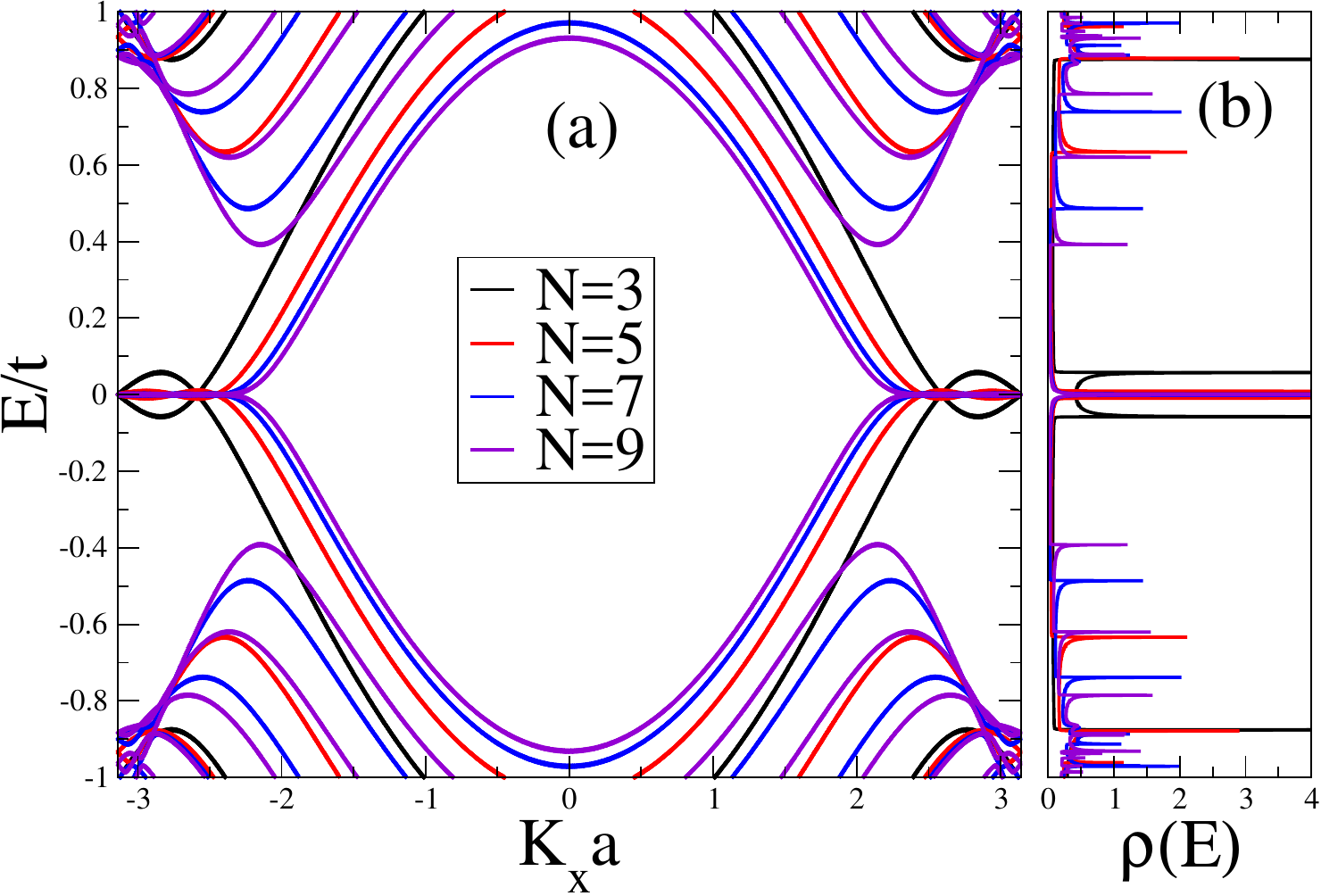} 
			\label{band_dos_impares}
    \end{subfigure}
		
			\vspace{0.5cm}
		\begin{subfigure}[tbh]{0.40\textwidth}
        \includegraphics[width=\linewidth]{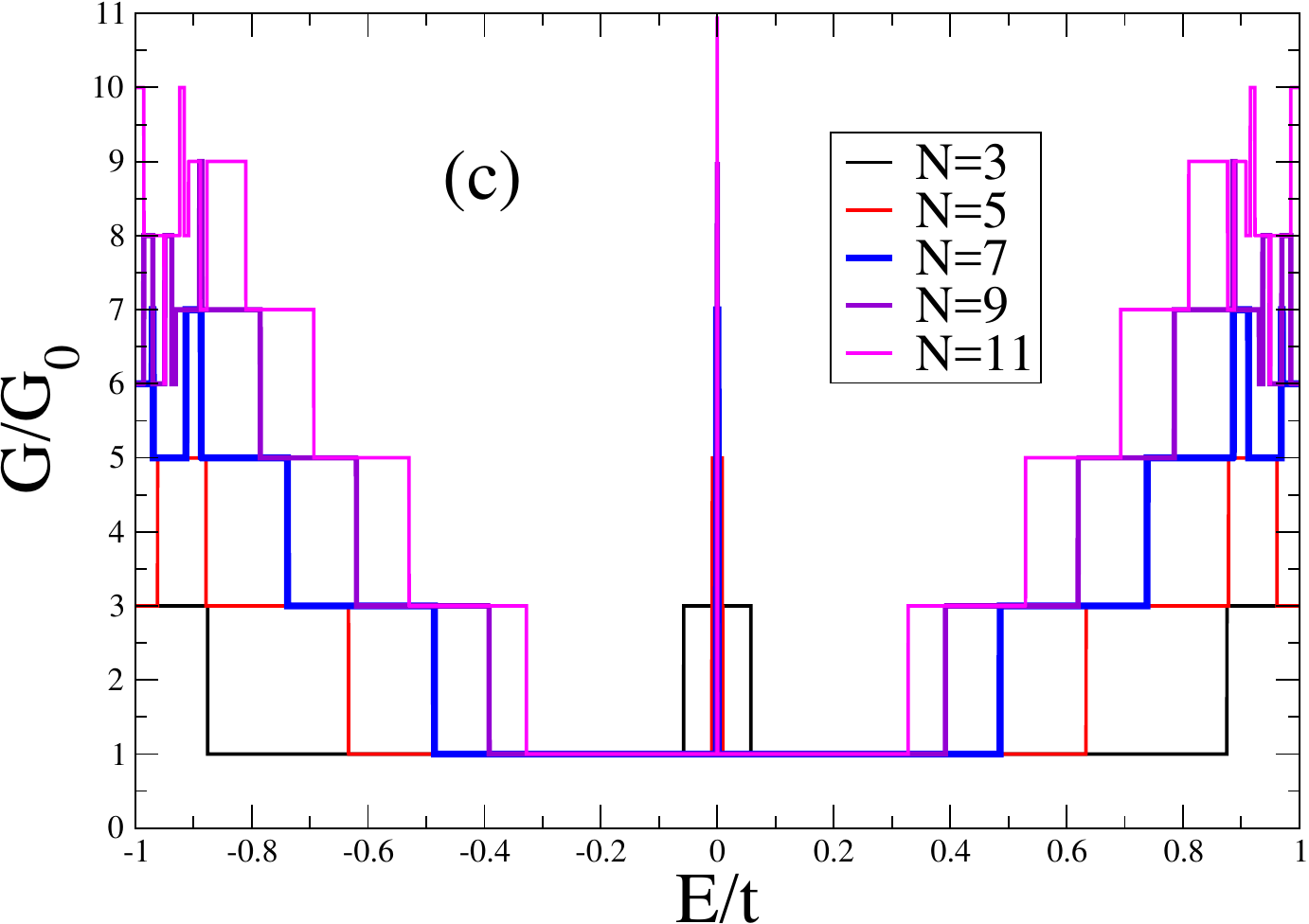} 
				\label{condutance_impar}
						\end{subfigure}
				\caption{(Color online) a) Band structure,  b) density of states and c) conductance of ZGNRs  with $t$ of the same sign as $t''$, and 
				$N=3,5,7,9,11$.}
					\label{band_dos_cond_2}											
\end{figure}

In Figs. (\ref{DFT_band_par}, \ref{DFT_band_impar})a,b we plot the band structure and the density of states, for even and odd ZGNRs, respectively. We only plot the branches close to the Fermi energy in order to compare to the  tight-binding results of the next section, as indicated in Figs. (\ref{band_dos_cond_1}, \ref{band_dos_cond_2})a; we set  the Fermi level to zero. The TB and LDA results are very similar, however, the LDA braiding results present some distortion as indicated in Fig. \ref{Zoom_par}. This happens because it  takes into account all the orbitals present in the ZGNRs, and the main cause of this particle-hole asymmetry is the $N2$ hopping, which is hiding inside the LDA calculations.

\section{Tight-binding calculations}
\label{sec4}

In order to investigate in more detail, the braiding of  edge states obtained in the last section by the LDA calculations, and the associated transport properties of narrow ZGNRs, we perform tight-binding calculations considering the effects of the NN and N3 hopping employing the tight-binding Hamiltonian 
( Eq. \ref{H2}). 

In the following calculations, we employ the realistic tight-binding parameters obtained in reference \cite{Tran17} that fit ribbon graphene structures to density functional theory (DFT) results: $t=2.8eV$, $t'=0.025t$, and $t''=0.15t$. It is interesting to observe that the absolute value of the N3 hopping is much larger than the N2, which reinforces our earlier supposition of only considering hopping that preserves chiral symmetry. In the next two subsections, for the calculation of the conductance we consider the central cell and the leads connected by the same NN hopping: 
$t_{c}=t$. Therefore, the two-terminal device of Fig. \ref{diagrama_condutancia}b reduces to an infinite ZGNR.

\subsection{N3 of the same sign as NN}
\label{subsec1}

In Figs. (\ref{band_dos_cond_1}, \ref{band_dos_cond_2})(a,b), we plot the band structure and the density of states, respectively, for even and odd ZGNRs, with $t$ of the same sign as $t''$. The net effect of the inclusion of N3 hopping is to lift the degeneracy of the edge states at the Fermi energy, producing the braiding of the valence and conduction bands \cite{Cristina2011,Korytar2014} and generating Dirac cones with non-commensurate wave vectors $\vec{k}$. For $N =2$, the Dirac cone moves from the border of the Brillouin zone (when $t''=0$) to a non-commensurate $\vec{k}$.  For even $N$, a gap is opened at the border of the Brillouin zone, while for odd $N$ this gap is closed, but in both cases with an increase in $N$ the density of states grows and generates a strong peak at the Fermi energy. A similar  effect occurs in the case of oligoacenes, formed by joining a finite number of benzene rings \cite{Korytar2014}. In this case, the gap shows an oscillatory behavior, depending on the number of benzene rings.

In the two insets of Fig. \ref{band_dos_cond_1}a, we represent the braiding of edge states (top inset) and the LDOS(E=0) at the Fermi energy (bottom inset), as a function of the transverse sites' position for a $6$-ZGNR. It is clear from the bottom inset that the LDOS assumes high values only at the border of the $6$-ZGNR and decays rapidly in a direction toward the center, where it vanishes. Therefore, the edge states' braiding produces a quantum wire that harbors a multi-channel conductance associated with the width $N$ of the ZGNR. The number of Dirac cones is $N/2$ for even $N$ and $(N+1)/2$ for odd $N$. When $N$ increases, the number of Dirac cones also increases, and the valence and conduction bands merge together, generating a metallic behavior for even or odd $N$, and the band structure tends to a case similar to only NN hopping being present, with the density of states exhibiting a strong peak at the Fermi energy. We think this is the reason why these braiding effects have not been receiving much attention up to now. The majority of the papers are focused on TB calculations for pristine graphene, large $N$ nanoribbons and in the context of the Kane-Mele \cite{Kane_Mele_05} generalized model \cite{Reich2002,Gunlycke08,Wu2010,Kundu11,Cristina2011,Hung13,Hung14,Chen15}, or more complete numerical LDA calculations for a $N=11$ zigzag ribon  \cite{Tran17}, in which the numerical resolution blurs the braiding effects of the edge states, which according to our calculations only occur for low $N$ ZGNRs. 

It is important to mention here that the edge braiding states are not affected by the intrinsic spin-orbit interaction (SO), since this interaction ($\lambda_{so}$) is negligible in graphene; $\lambda_{so}=1.3\mu eV$ \cite{Liu2011}. However, the 2D materials, such as silicene \cite{Paola_silicene}, germanene \cite{Zhang2016}, and stanene \cite{Zhu2015} exhibit a much stronger intrinsic SO interaction than graphene \cite{Martins17}, which is due to their higher atomic number and planar buckling structures. In silicene, the SO is very low; $\lambda_{so}=3.9 meV$, and can also be neglected. However, in germanene: $\lambda_{so}=46.3 meV$ and in stanene: $\lambda_{so}=64.4 meV$ \cite{Liu2011}, this interaction is higher and the degeneracy of the edge braiding states is lifted by the SO interaction, and a tiny gap of the order of $meV$ is opened in the density of states. 

Another important result of the present paper is obtained as a consequence of the Stoner criterion, which states that the magnetic order is favored  if $U\rho(E_{f})>1$ \cite{Ziman72,Teodorescu008}, where $U$ is the electronic local Coulomb correlation and $\rho(E_{f})$ is the non-interacting density of states at the Fermi energy. Considering a NN tight-binding calculation, the ZGNRs edge states were theoretically predicted \cite{Nakada96} to couple ferromagnetically along the edges and antiferromagnetically between them. When only NN neighbors hopping are taken into account, the Stoner criterion is always satisfied, due to the moderate electronic correlation of the order of $U \simeq 0.8t$ present in those ZGNRs, \cite{Jung09} and the strong peak at the Fermi energy.  However, when the N3 hopping is included in the calculations, this strong peak disappears for low-width ZGNRs as indicated in  Figs. (\ref{band_dos_cond_1}, \ref{band_dos_cond_2})b, and the Stoner criterion could not be satisfied. According to the density of states curves represented in Figs. (\ref{band_dos_cond_1}, \ref{band_dos_cond_2})b, for 
$N=(2,3)$,  $\rho(E_{f})=(0.20,0.40)$, and to satisfy the Stoner criterion, we must have 
$U>5.0t,U>2.5t$. For $N=(4,5)$, $\rho(E_{f})=(0.83,1.70)$, the Stoner criterion is only satisfied if $(U>1.20t,U>0.59t)$, respectively. Therefore, according to our TB  calculations, low-width ZGNRs with $N \lessapprox 4$ could not develop magnetic order at their edges. 

In Figs. (\ref{band_dos_cond_1}, \ref{band_dos_cond_2})c, we plot the conductance in units of the quantum conductance $G_{o} = 2e^{2}/h$ for even and odd ZGNRs, respectively. Due to the presence of the edge states' braiding, new conductance channels are open at integer steps, and the conductance at the Fermi energy assumes integer multiples of $G_{o}$: for $N=(2,3,4,\cdots)$, $G=(2,3,4,\cdots)G_{o}$, respectively. These ZGNRs behave as a multiple conductance quantum wire that works as a ``current filter". The conductance value at the Fermi energy attains high integer values that can be controlled by their width $N$. Therefore, the system exhibits a potential to be employed in technological applications. Another interesting aspect of the conductance results is the fulfillment of the electron-hole symmetry in relation to the Fermi energy, which is a consequence of the chiral symmetry of the Hamiltonian, as discussed in the introduction of this paper. 

Another interesting point of the present results is  the values assumed by the density of states at the Fermi level, obtained from the unplolarized LDA calculations,  as plotted in Figs. (\ref{DFT_band_par}, \ref{DFT_band_impar})b. For $N=2$, $\rho(0)$ assumes a value close to zero, and for $N=3$ it is of the order of the unit, but increases with the increasing of $N$, which agrees with our TB calculations as indicated in Figs. (\ref{band_dos_cond_1}, \ref{band_dos_cond_2})b.

\subsection{N3 and NN of opposite signs}

Contrary to the case studied in the earlier subsection, we did not identify any known $2D$ real system with the N3 and NN hopping having opposite signs. However, a recent paper described an experimental realization of tunable optical sawtooth and zigzag lattices employing optical lattices of ultracold atoms of alkali-earth-like bosons and fermions (${}^{(173,174)}$Yb and 
${}^{(84,85)}$Sr isotopes) \cite{Zhang2015}. The authors attain precise control over the intra and inter-unit-cell hopping. Therefore, the results presented in this subsection represent theoretical possibilities \cite{Kivelson2013,Cristina2011} that we expect to be realized employing optical lattices. 

\begin{figure}
     \begin{subfigure}[t]{0.45\textwidth}
        \includegraphics[width=\linewidth]{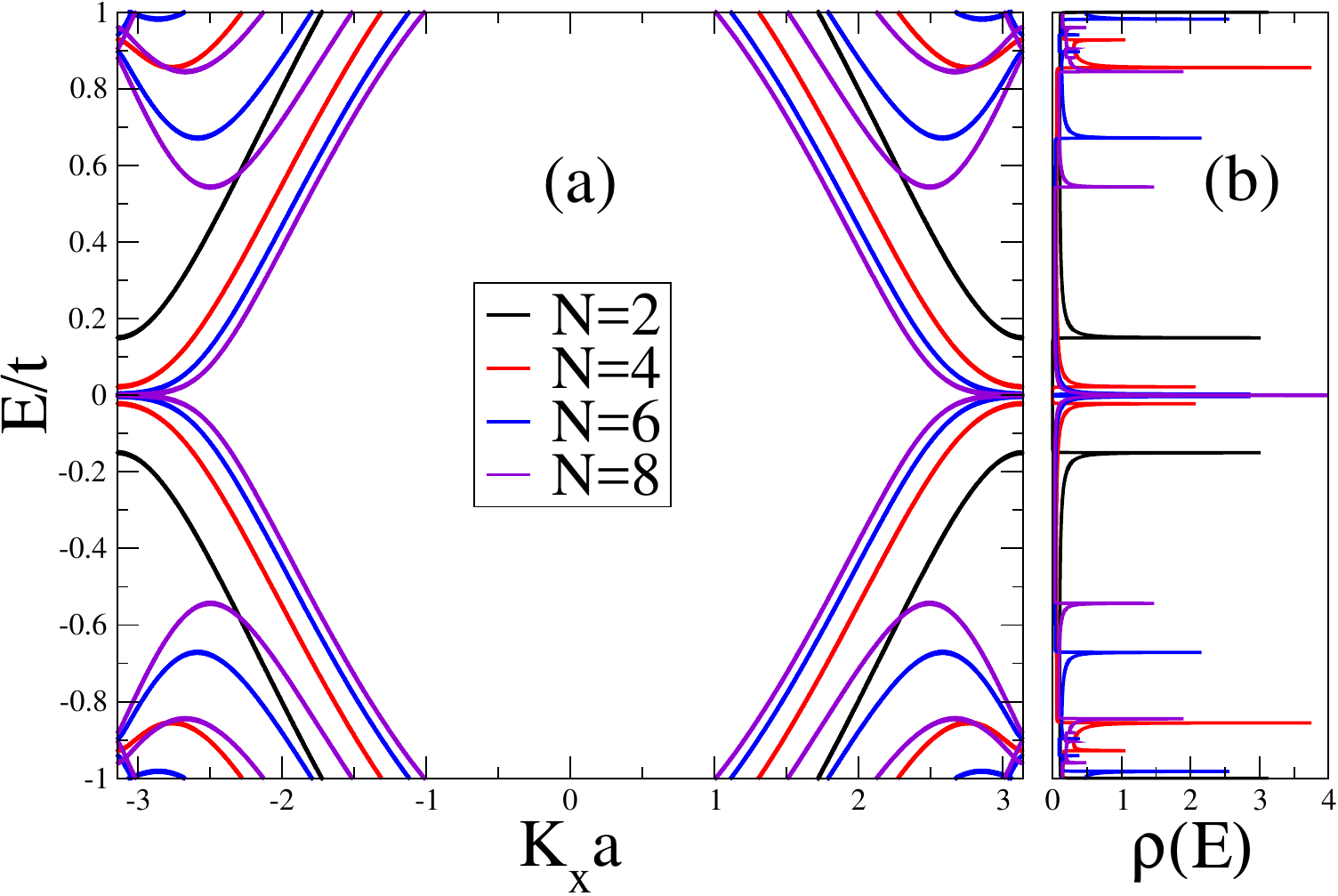} 
\label{band_dos_positive_par}
    \end{subfigure}

    \vspace{0.6cm}
    \begin{subfigure}[t]{0.45\textwidth}
        \includegraphics[width=\linewidth]{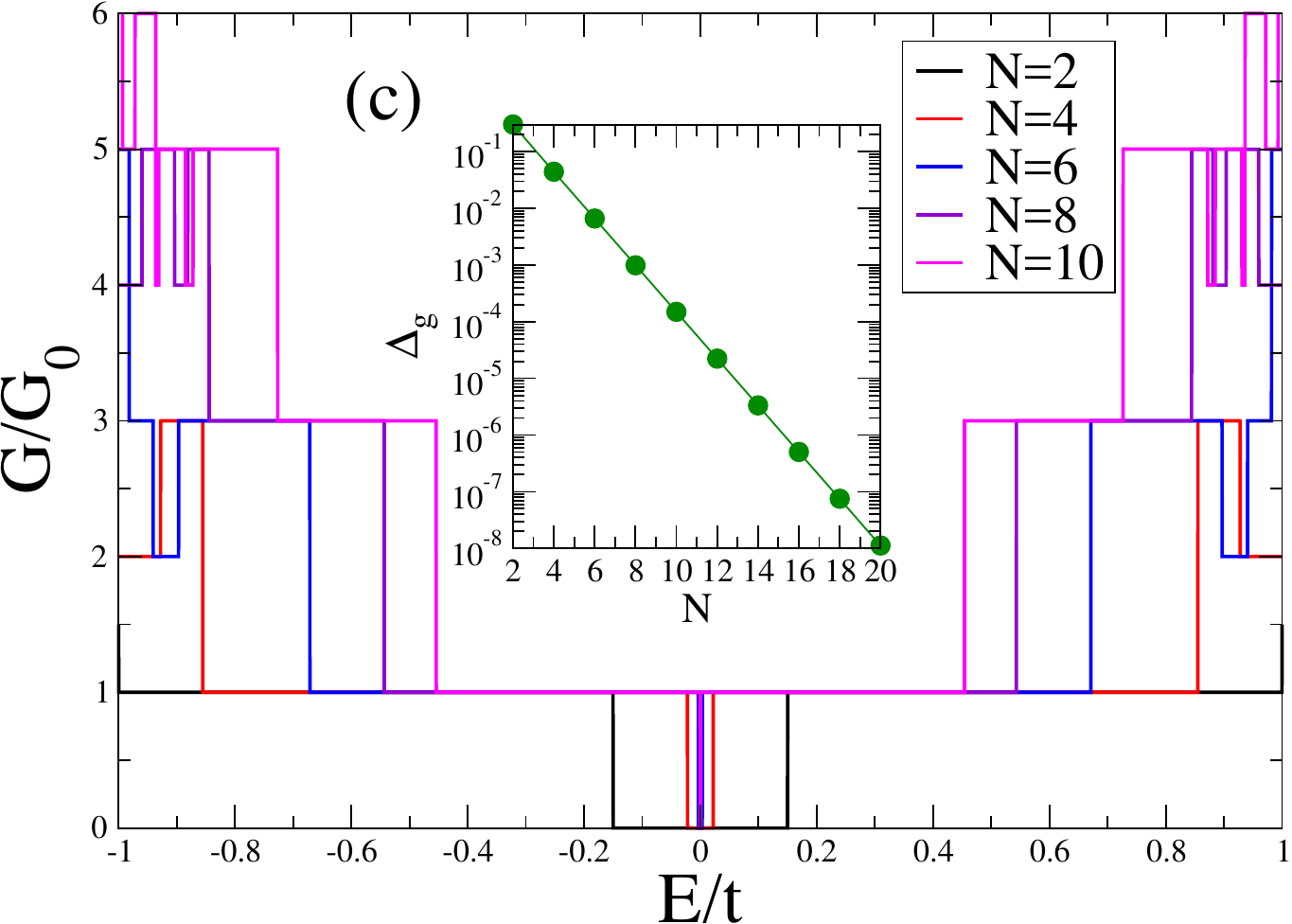}          
				\label{condutance_par_3}
						\end{subfigure}
				\caption{(Color online) a) Band structure,  b) density of states and c) conductance of ZGNRs with $t$ and $t''$ of opposite signs, and $N=2,4,6,8,10$.}
		\label{band_dos_cond_3}
\end{figure}
\begin{figure}
 \begin{subfigure}[t]{0.45\textwidth}
        \includegraphics[width=\linewidth]{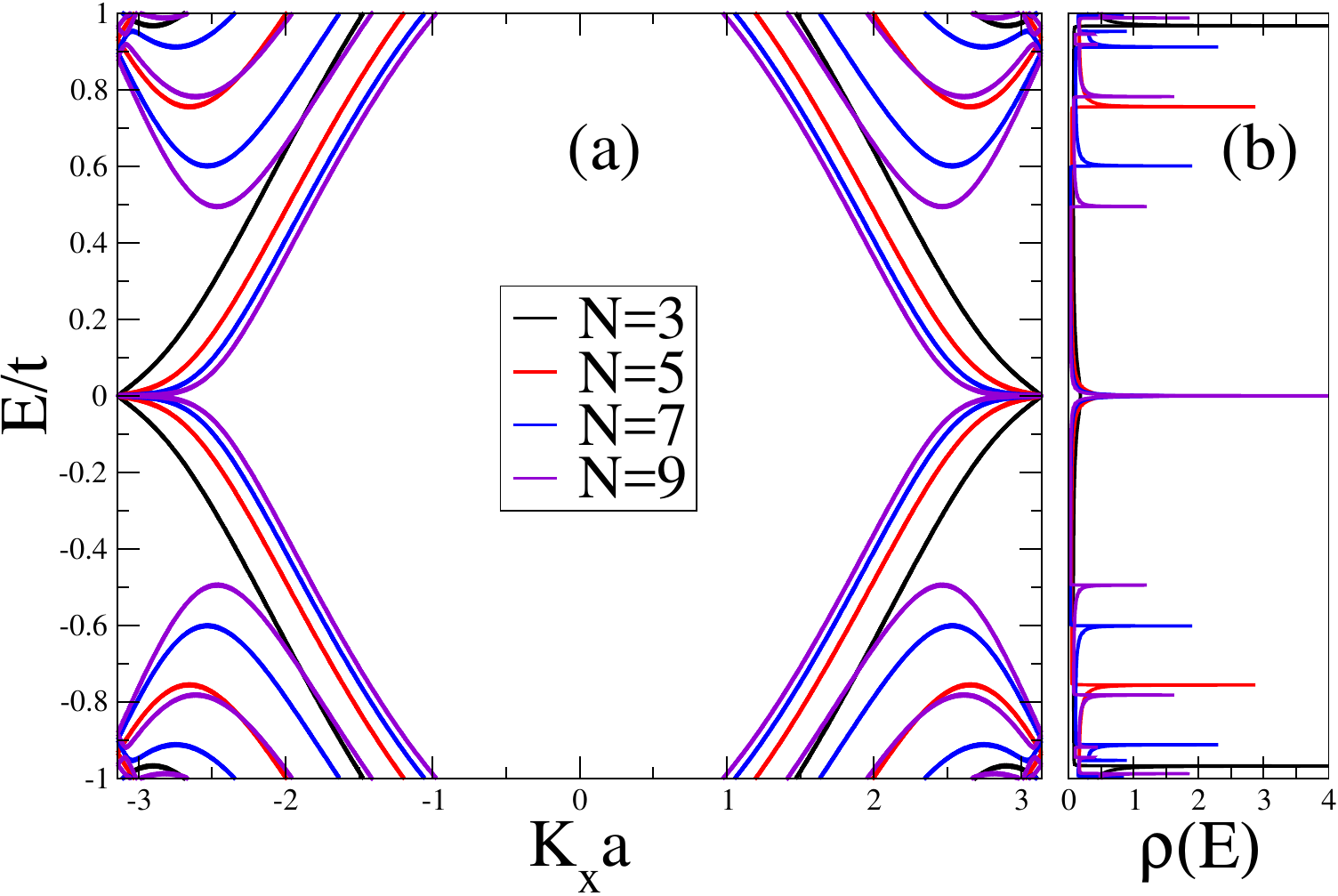} 
\label{band_dos_positive_par}
    \end{subfigure}

    \vspace{0.6cm}
    \begin{subfigure}[t]{0.45\textwidth}
         \includegraphics[width=\linewidth]{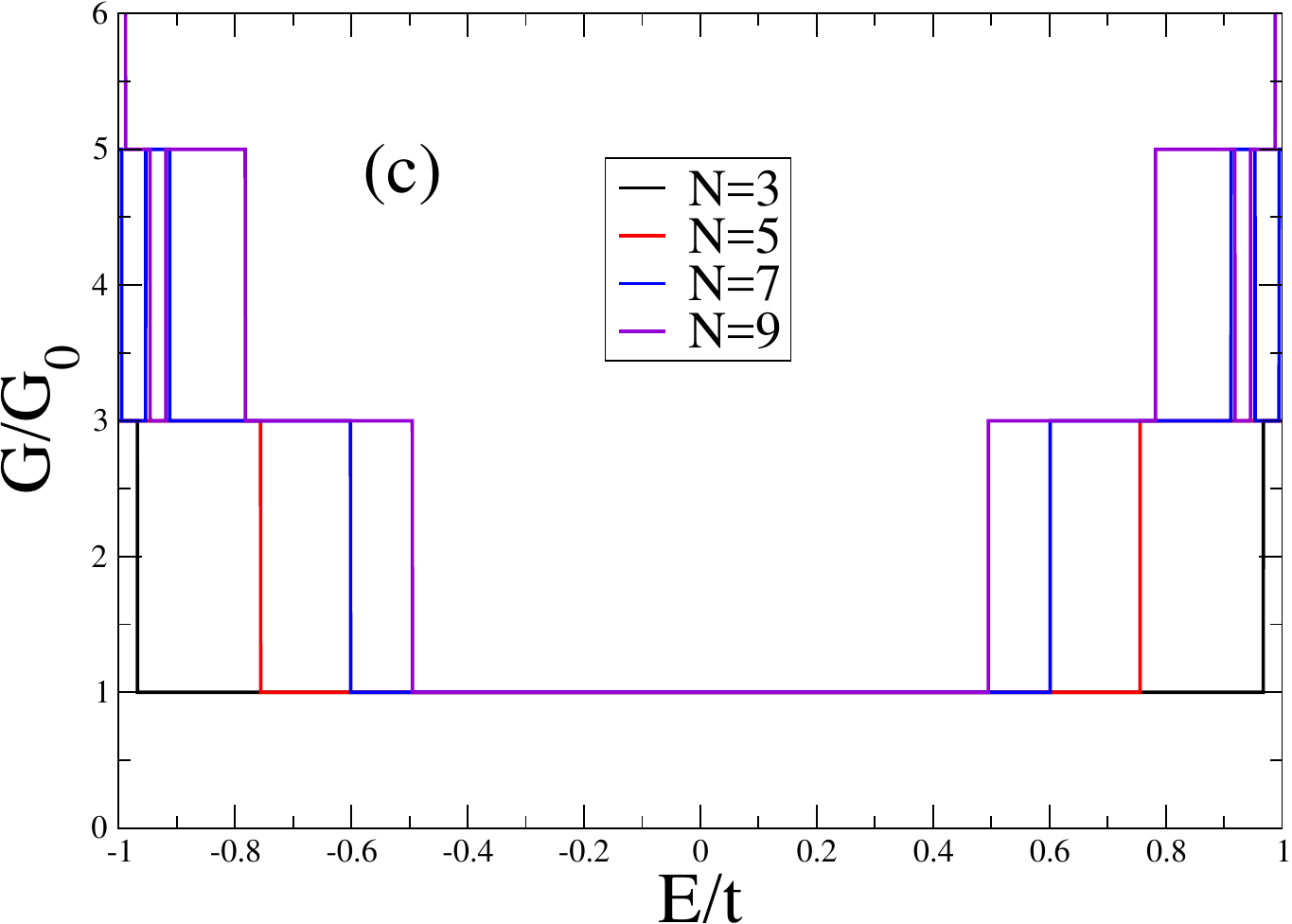}      
				\label{condutance_par_4}
						\end{subfigure}
\caption{(Color online) a) Band structure,  b) density of states and c) conductance of ZGNRs with $t$ and $t''$ of opposite signs, and 
$N=3,5,7,9$.}
    		\label{band_dos_cond_4}
\end{figure}
In Figs. (\ref{band_dos_cond_3}, \ref{band_dos_cond_4})(a,b), we plot the band structure and the density of states for even and odd ZGNRs, respectively, considering $t$ and $t''$ of opposite signs. For low, even $N$ values, the system behaves as a band insulator, with the band structure exhibiting a gap $\Delta_{g}$ at the Fermi energy, but with an increase in $N$, this gap decreases logarithmically, as indicated in the inset of Fig. \ref{band_dos_cond_3}c, and tends asymptotically to zero, leading to the system's undergoing an insulator-metal transition. This is a Lifshitz-type transition \cite{Lifshitz60}, but here the topology of the Fermi surface changes in a discrete way. For the odd $N$ case, as indicated in Fig. \ref{band_dos_cond_4}(a,b), the valence and conduction bands cross each other at the borders of the Brillouin zone, and the system is always metallic.

In Figs. (\ref{band_dos_cond_3},\ref{band_dos_cond_4})c, we plot the conductance for even and odd ZGNRs in units of  $G_{o} = 2e^{2}/h$,  respectively. For even $N$, the ZGNRs behave like an insulator, with the gap controlled by the width of the nanoribbon. The gap also decreases logarithmically with an increase in the nanoribbon width $N$, as indicated in the inset of Fig. \ref{band_dos_cond_3}c, and tends asymptotically to a semimetal band.  This kind of control is quantified by the possibility of switching between different states of electrical conductivity of the material. The ratio of the ON- and OFF-state conductance of field-effect-transistors, probed at zero gate bias and at low drain bias, is defined as the ON/OFF ratio of the material \cite{Xia10,Yung13}. In the case of graphene, it is very low when compared to silicon, i.e. graphene continues to conduct a lot of electrons even in its OFF state. One possible candidate for overcoming the limitations of graphene is the ZGNRs discussed here, whose gap decreases logarithmically with an increase in $N$, allowing a fine control over the conductance of charge carriers and making feasible the adjustment of the variation of the ON/OFF ratio to a value adequate to the technological needs. For odd $N$, the conductance exhibits a behavior similar to the metallic armchair nanoribbons, with $N_{a} = 3p -1$, where $p$ is an integer number \cite{katsunori}.

\begin{figure}[tbh]
\begin{center}
\includegraphics[clip,width=0.45\textwidth,angle=0.0]
{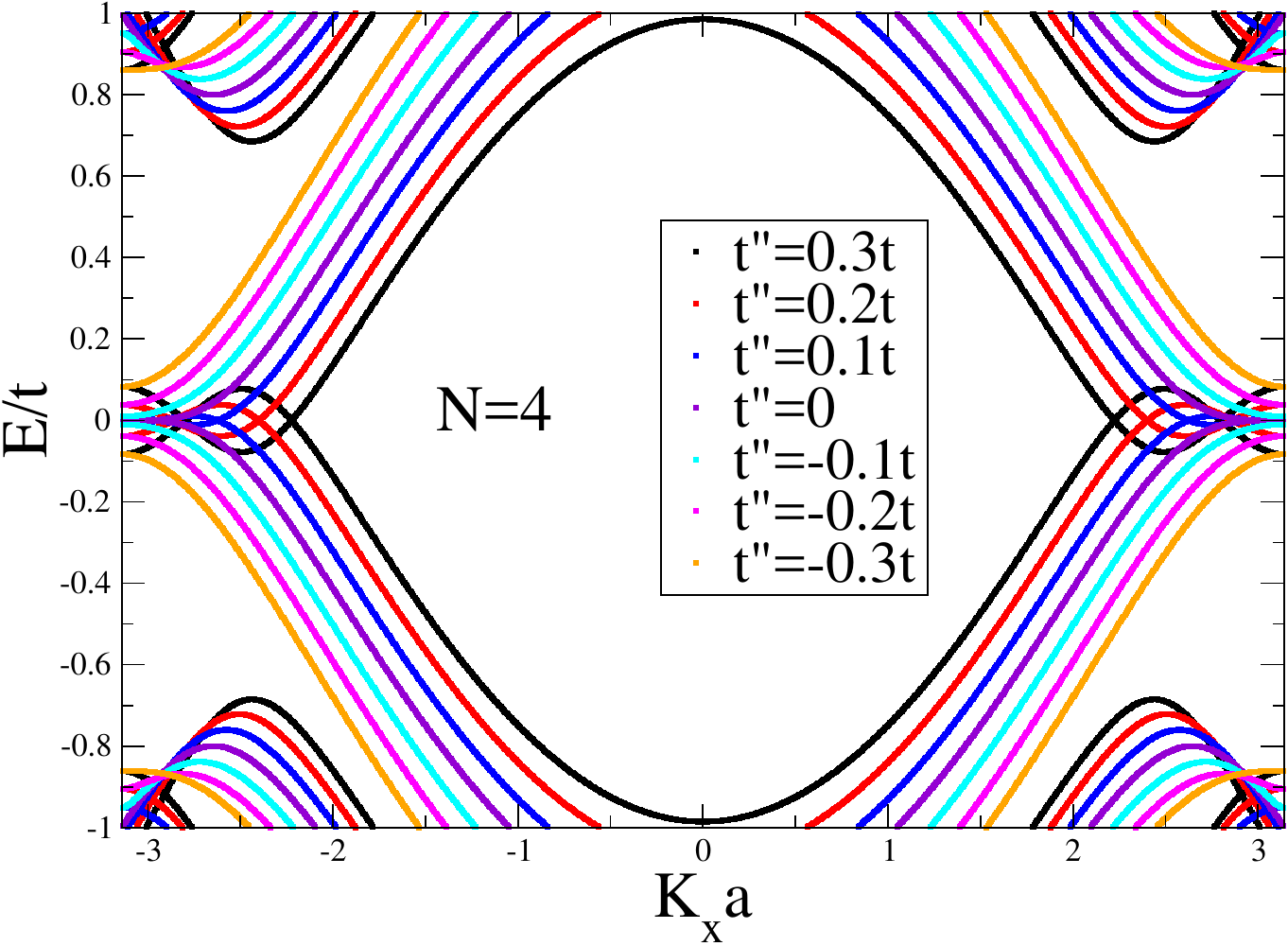}
\end{center}
\caption{(Color online) Band structure of a $4$-ZGNR  considering different N3 hopping: $t''=(-0.3,-0.2,-0.1,0,0.1,0.2,0.3)t$.}
\label{Lifshitz}
\end{figure}
\begin{figure}
    \begin{subfigure}[!]{0.45\textwidth}
        \includegraphics[width=\linewidth]{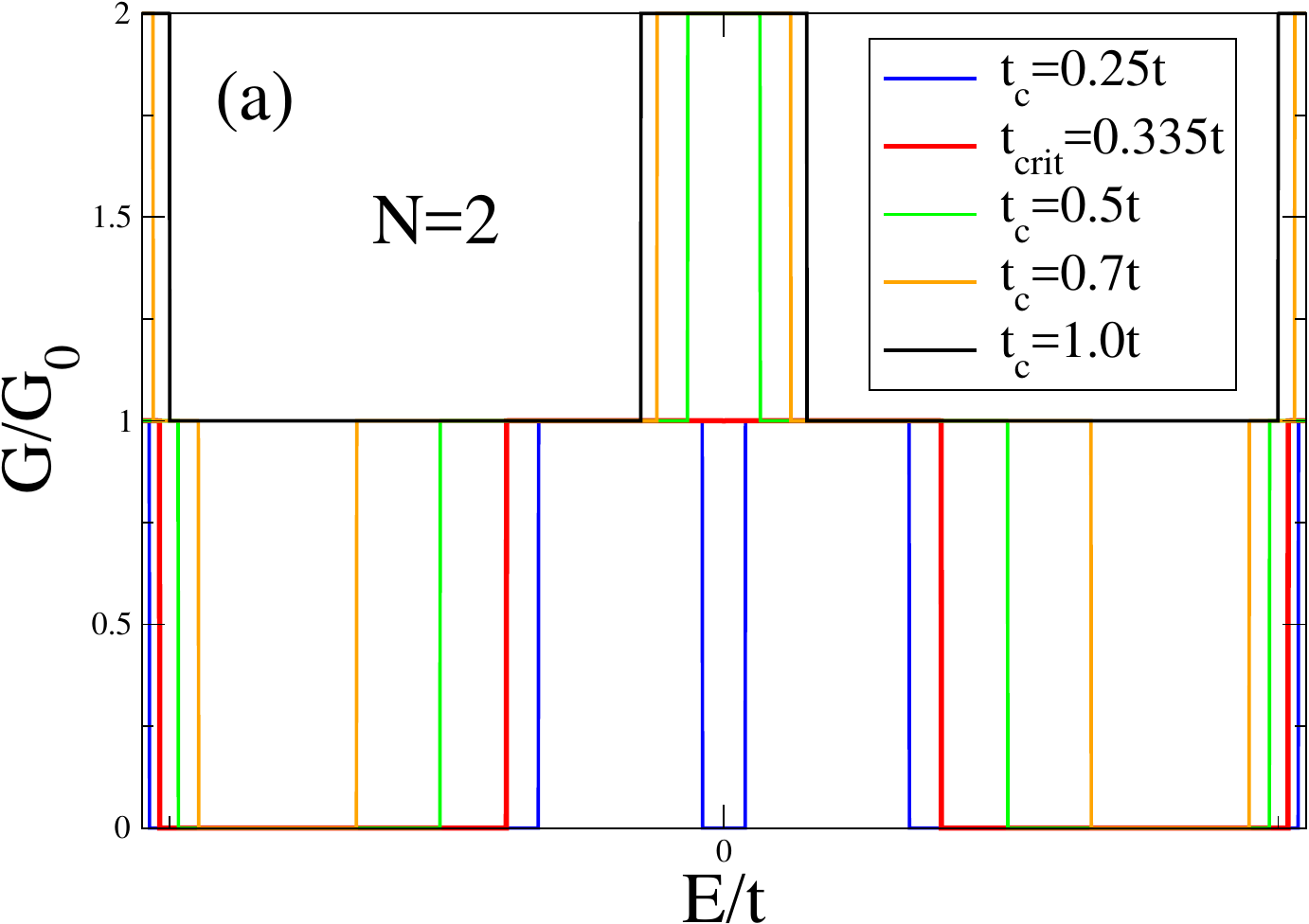} 
 \label{Device1}
    \end{subfigure}

    \vspace{0.55cm}
    \begin{subfigure}[!]{0.45\textwidth}
        \includegraphics[width=\linewidth]{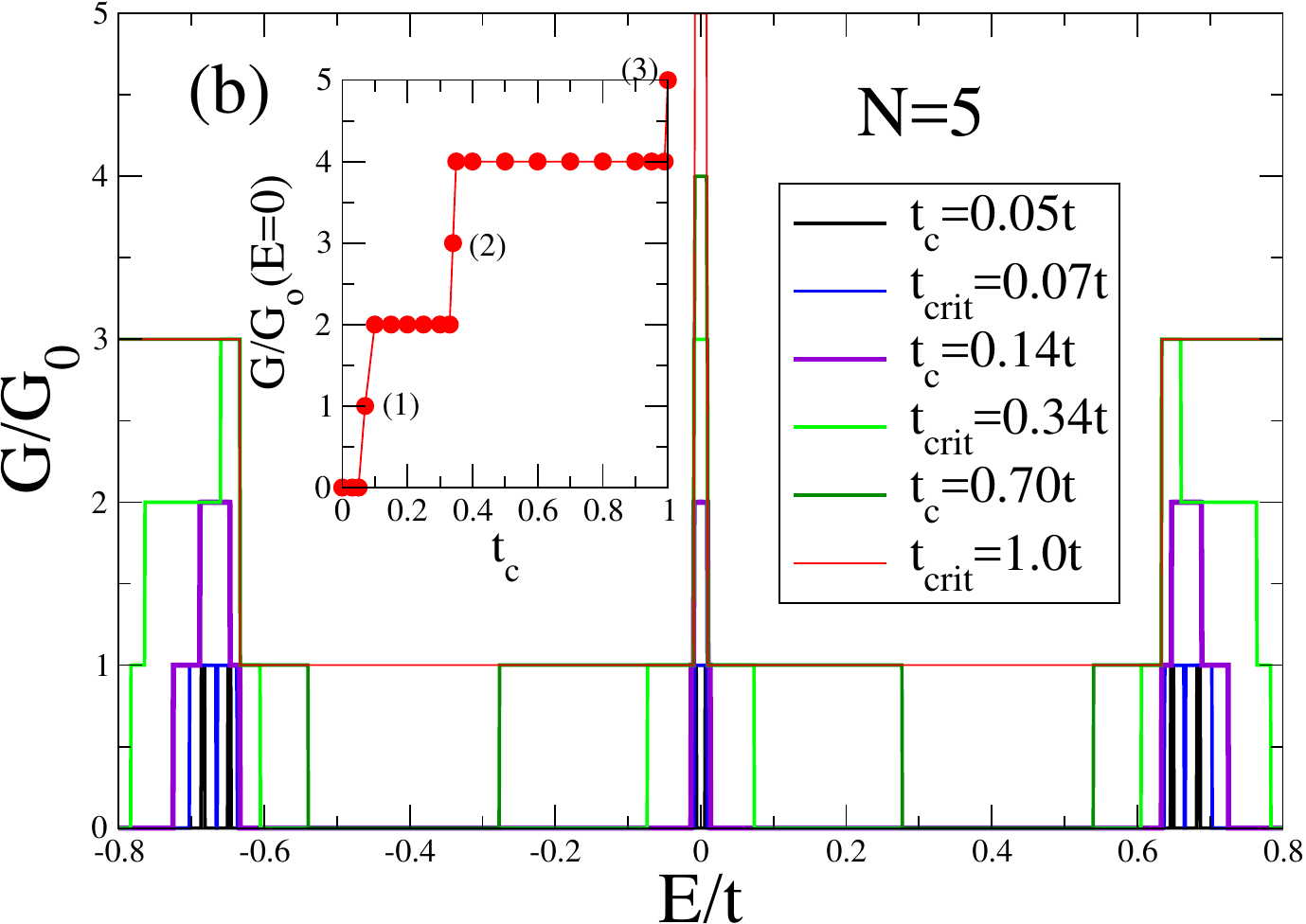} 
								\label{Device2}
						\end{subfigure}
\caption{(Color online) a) Two terminal ZGNR device conductance for  $N=2$, and for b) $N=5$, considering different couplings $t_{c}$ connecting the central cell (indicated by a green box in Fig. \ref{diagrama_condutancia}b) to the leads.}
		\label{Device12}
\end{figure}
In Fig. \ref{Lifshitz}, we plot the band structure of  a $4$-ZGNR considering  the hopping $t''$ variable. For negative $t''$ values, the 
$4$-ZGNR behave as a band insulator, and for positive $t''$ values, it becomes  metallic. Due to the variation of the N3 hopping, the nanoribbon evolves continuously from a gap situation to a metallic Dirac multi-channel quantum wire, undergoing a Lifshitz \cite{Lifshitz60} insulator-metal transition at the quantum critical point $t''=0$. This kind of phase transition was reported in a honeycomb two-leg ladder \cite{Kivelson2013} for $N=2$. 

\subsection{An application: a quantum dot connected do semi-infinite leads}

In this subsection, we consider in all the calculations the central cell indicated by a green box in Fig. \ref{diagrama_condutancia}b, connected to the  leads by a variable hopping $t_{c} \leq t$. The hopping $t_{c}$  can be tuned by voltage gates.

The absence of a true gap in graphene sheets constitutes a great problem from the technological point of view. In order to employ this material as a substitute of silicon for use in development of logic transistors, a fine control over the conductance of charge carries is necessary.   In this subsection we show that the system can be tuned to an insulator-metal transition, with the conductance at the Fermi energy exhibiting a staircase behavior as a function of the hopping $t_{c}$.

In Figs. \ref{Device12}(a,b), we plot the conductance of the two-terminal device, represented in Fig. \ref{diagrama_condutancia}b, as a function of the energy $E$, for two ZGNRs with $N=2$ and $N=5$, respectively. Varying the hopping $t_{c}$ in the interval $[0,1]t$, the system can be tuned to an insulator-metal transition, as indicated in Fig. \ref{Device12}a for $N=2$, where the transition occurs at a critical hopping $t_{crit}=0.335t$. When $N=5$, as plotted in Fig. \ref{Device12}b, this transition occurs at $t_{crit}=0.07t$, with the inset of the Fig. \ref{Device12}b showing that the conductance at the Fermi energy exhibits a staircase behavior as a function of $t_{c}$, increasing $2G_{o}$ in each step. In the odd conductance channels, there is only one critical point that is formed by the closing of a step, followed by the opening of another conductance channel that increases the conductance by one more $G_{o}$ unit. We represent those points by the numbers: 
$(1)$ corresponding to the first transition ($t_{c}=0.07t$) at $1G_{o}$, $(2)$ corresponding to the second transition ($t_{c}=0.34t$) at 
$3G_{o}$, and $(3)$ corresponding to the third transition ($t_{c}=1.0t$) at $5G_{o}$. Therefore, the system is tuned through all the possible integer conductance units.

\section{Polarized density functional theory (LSDA)  simulations}
\label{sec5}

In order to investigate the magnetic nature of the fundamental state of the low-width ZGNRs, we return to the DFT calculations, but now employing the spin polarized version of this theory (LSDA), as it is implemented in the SIESTA package \cite{SIESTA}.

\begin{figure}[tbh]
\begin{center}
\includegraphics[clip,width=0.45\textwidth,angle=0.0]
{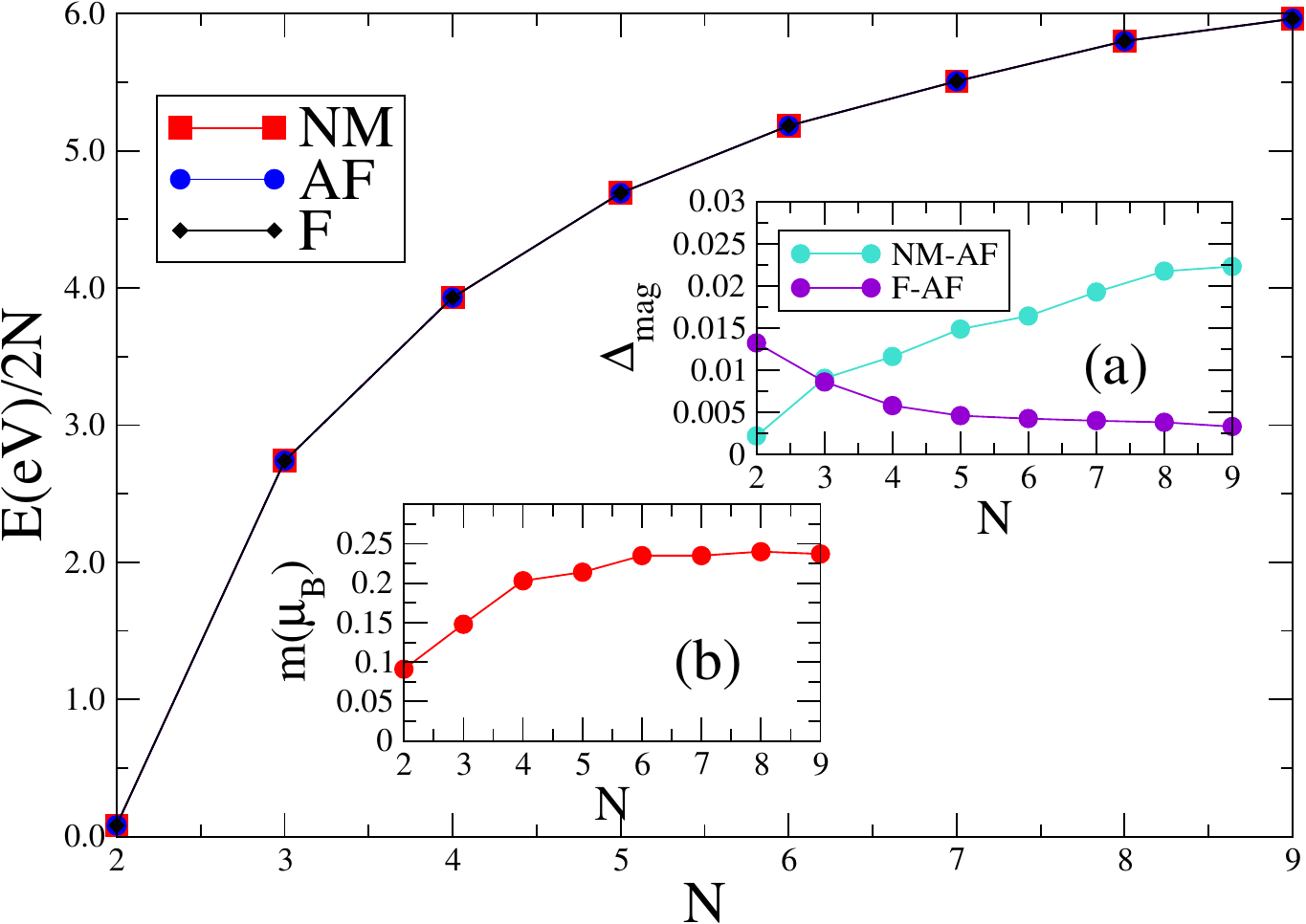}
\end{center}
\caption{Energy in electron Volts units $E(eV)/2N$ per site as a function of the low-width ZGNRs width $N$, corresponding to the NM,  AF and F  configurations between the edges.}
\label{PAFF}
\end{figure}

The Lieb theorem  \cite{Lieb89} establishes that the bipartite lattice, present in pristine graphene and described by the Hubbard hamiltonian in the half filling case, must have the total magnetic moment $J$ of the ground state null, such that 

\begin{equation}
J={1\over 2}||B|-|A|| .
\label{Total_moment}
\end{equation}
where $B$ and $A$ represent the graphene sublattices, with the number of sites given by 
$|A|$ and $|B|$, respectively. This result must be fulfilled by our tight-binding hamiltonian defined by Eq. \ref{H2}, that exhibits chiral symmetry, once we take into account only $NN$ and $N3$ hopping. However, this symmetry is not present in the DFT simulations, since this method takes into account all the orbitals present in the ZGNRs and, consequently, the processes associated with the hopping $N2$ that break particle-hole symmetry of the Hamiltonian. In this way, the Lieb theorem is not valid, and the ZGNRs could develop any magnetic order or no magnetic order at all. To investigate the possible magnetic ground state of those nanoribbons, we calculate the total energy for each of them when there are magnetic moments along the edges for non-magnetic (NM), ferromagnetic (F) and antiferromagnetic (AF) configurations. The results are presented in Fig. \ref{PAFF}

\begin{figure}
    \begin{subfigure}[!]{0.45\textwidth}
        \includegraphics[width=\linewidth]{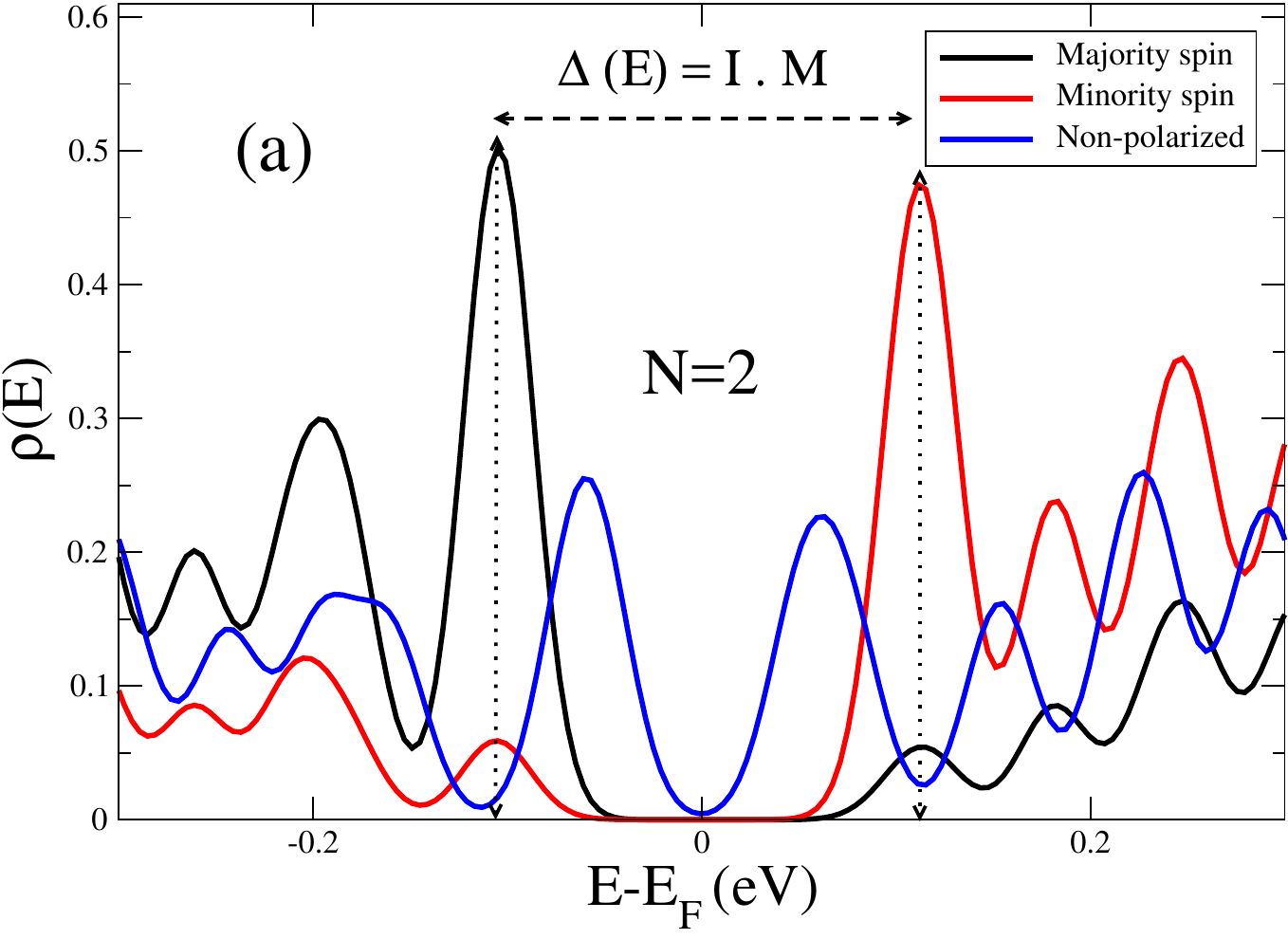} 
 \label{Polarized_2}
    \end{subfigure}

    \vspace{0.55cm}
    \begin{subfigure}[!]{0.45\textwidth}
        \includegraphics[width=\linewidth]{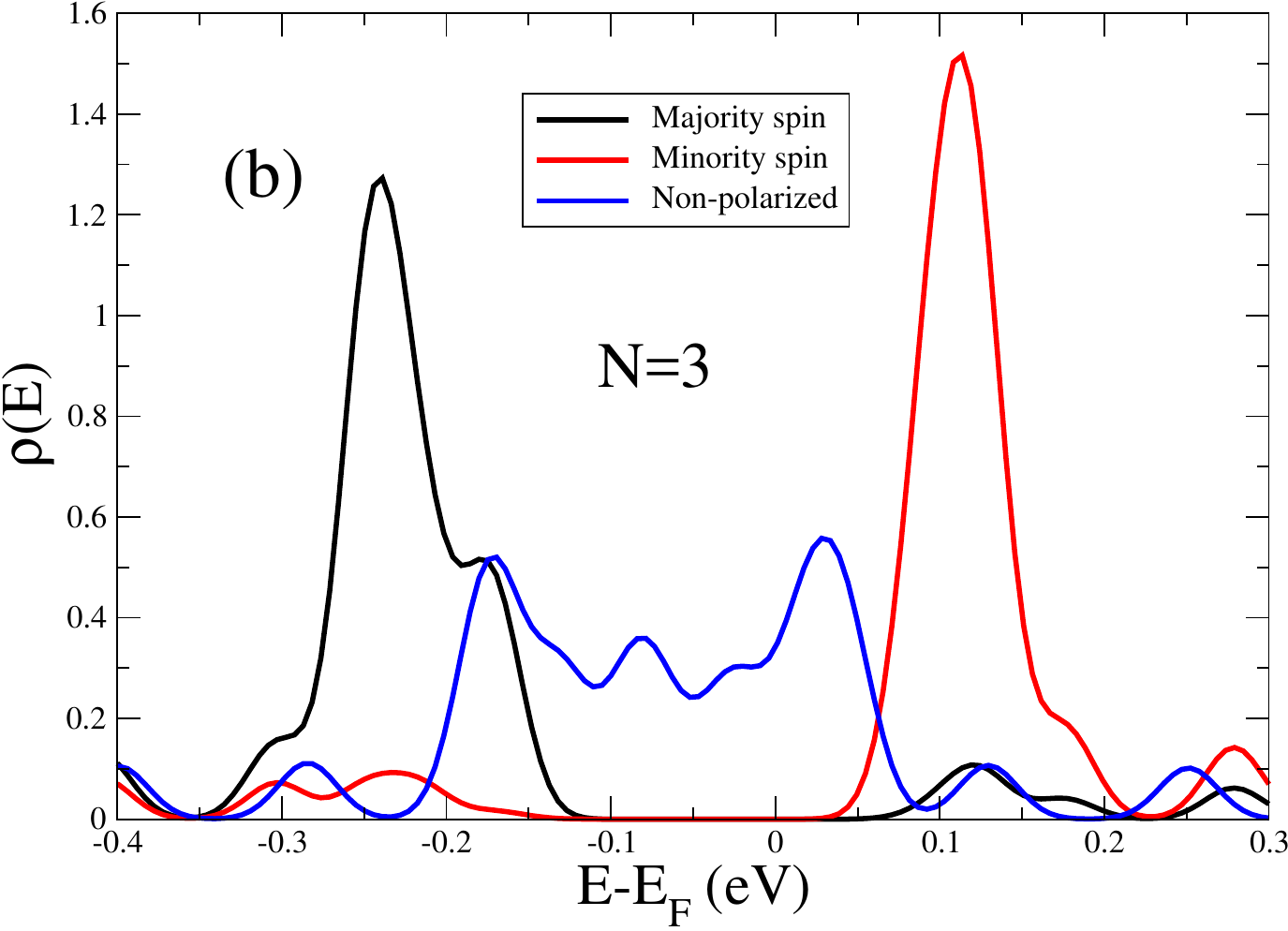} 
\label{Polarized_3}
\end{subfigure}
\caption{(Color online) Polarized density of states obtained from the spin-unrestricted LSDA calculations for a) $N=2$ and b) $N=3$.}
\label{Polarized}
\end{figure}

In  Fig. \ref{PAFF}, we plot the total energy $E(eV)/2N$ per site (there are $2N$ sites along the ZGNRs transverse direction) as a function of $N$, corresponding to the NM, AF and F configurations between the edges. The magnetic energy difference $\Delta_{mag.}$ between the curves is very small and attains a flat regime as N increases, meaning that correlations between edges become less important as the ZGNRs width becomes larger. 

The inset (a) shows $\Delta_{mag}$ per unit cell corresponding to the energy difference between ferro and antiferromagnetic ($F-AF$) configurations as a function of $N$, whereas the energy difference between the non-magnetic and antiferromagnetic ($NM-AF$) cases is given with respect to the edge atoms. The AF configuration wins in all cases, although we still believe that for lower  $N$ ZGNRs we need more accurate calculations. 

Using standard Mulliken population analysis in the inset (b), we show the magnetic moment 
$m(\mu_{B})$ in Bohr magneton units associated to the edge atoms; it decreases as the number of zigzag lines does and attains its minimum value for $N=2$, where the competition occurs between NM and AF configurations, but it does not vanish as we expect from the non-satisfaction of the previously discussed Stoner criterion. On the other hand, a recent work \cite{Caio14}, which studied ZGNRs employing a TB Hubbard model and considering $NN$ and $N2$ hopping, obtained a vanishing magnetization for low N, even satisfying the Stoner criterion. 

In Fig. \ref{Polarized} we plot the polarized density of states corresponding to different spin orientations, obtained from the  unrestricted spin-polarized LSDA  corresponding to a) $N=2$ and b) $N=3$. We also plot the non-polarized solution obtained from the LDA. The non-polarized density of states in both cases is very low at the Fermi level, which will lead to the non satisfaction of the Stoner criterion.

Concerning the calculation of the Stoner criterion employing LSDA, it is worth to mention that the electronic correlations were included in this theory through a weak-coupling mean-field theory. LSDA assumes that the electronic correlation, like the Hubbard parameter $U$, is small when compared to the bandwidth $W$ and, as a consequence of this correlation, a gap is opened at the Fermi energy. For that purpose, we followed the method given by \cite{Zeller06}, in which the exchange integral (=Stoner parameter  $I$) is calculated as an energy difference, once we take into account the polarized LDOS for both majority and minority spin populations, as indicated by the horizontal arrow in Fig. \ref{Polarized}a. This energy difference can be expressed in terms of the local magnetic moment  $m$, which is obtained through an integral of the magnetization density M(r) over the atomic unit cell $\Omega$
\begin{equation}
m=\int_{\Omega} M(r) dr ,
\label{momento}
\end{equation}
which was also calculated and shown in the inset b) of the Fig. \ref{PAFF}. The energy difference $\Delta$ is calculated from  Figs.  \ref{Polarized}(a,b), as 
\begin{equation}
\Delta (E)= m \cdot I
\label{Stoner1},
\end{equation}
and the Stoner criterion holds as long as

\begin{equation}
\rho_{o}(E_F)\cdot I>1,
\label{Stoner2}
\end{equation}
where $\rho_{o}(E_{F})$ is the value of the LDOS at the Fermi level obtained from the  non-polarized LDA calculation in Figs.  \ref{DFT_band_par} and \ref{DFT_band_impar}  and reploted in Figs.  \ref{Polarized}(a,b) . 

The Stoner parameter $I$ is calculated from the direct application of Eq. \ref{Stoner1} and the results are:  
 $I=2.37 eV=0.85t$ for $N=(2,3)$ (the same result of $I$ for both, $N=2,3$ is fortuitous)  and $I=2.12 eV=0.76t$ for $N=4$. Now extracting the unpolarized density of states, at the Fermi level  $\rho_{o}(E_{F})$, from  Figs. \ref{Polarized}(a,b), and applying Eq. \ref{Stoner2} we obtain for $N=2$, $I\rho(E_{F})=0.10$; for $N=3$,  $I\rho(E_{F})=0.76$ and for $N=4$, $I\rho(E_{F})=2.12$. 
 
Therefore, the results employing TB or LSDA agree in predicting that  narrow-width ZGNRs with $N=(2,3)$ do not satisfy the Stoner criterion,  and for $N=4$, it is satisfied but has the same order of magnitude of the $N=(2,3)$ case. This is a robust result, since the density of states, at the Fermi level $\rho_{o}(E_{F})$ of these nanoribbons, is very low in both methods. However, according to the inset (b) of Fig. \ref{PAFF}, the ground state of these ZGNRs is antiferromagnetic: the Stoner criterion fails but the system exhibits magnetic order, which is a contradiction. As such, we conclude that even LSDA \cite{SIESTA} is not able to resolve the magnetic order of these narrow-width ZGNRs. 

The electronic correlations associated with the Stoner parameter $I$ present the same order of magnitude of the parameter $U$, employed in the literature \cite{Tran17}: $I=0.85t$ for $N=(2,3)$ and $I=0.76t$ for $N=4$. Therefore, further calculations, like LDA+U method could not resolve this issue. The antiferromagnetic ground state should be a consequence of the mean-field character of the LSDA, and one possibility to go further is to employ the GW approximation, which includes a screened-exchange self energy as a result of the renormalization of the bare Coulomb interaction. Besides, the system could not be robust under the quantum fluctuations present in the edge magnetism of ZGNRs \cite{Wessel17}, and a method that includes local quantum fluctuations, like DMFT combined with GW approximation \cite{Tomczak2017} could contribute to elucidate the real magnetic nature of the ground state of these narrow zigzag nanoribbons.

\section{Conclusions}
\label{sec6}
 
In this paper, we performed LDA simulations and  tight-binding calculations on the narrow width ZGNRs, that confirm the braiding of the valence and conduction bands. We developed  a tight-binding study of low-width ZGNRs, taking into account the nearest-neighbor hopping NN and the third-neighbor hopping N3.  We calculated the band structure, the density of states, and a two-terminal device conductance employing the Landauer-Buttiker formalism. In order to investigate the magnetic nature of the fundamental state of these nanoribbons, we  employed  the spin polarized version of the density functional theory (LSDA).

Within the TB method, we analyzed two situations: when the NN and N3 hopping have the same sign and when they have opposite signs. In the first case, the ZGNRs always exhibit a metallic behavior, as a consequence of the braiding of the branches of the conduction and valence bands. New conductance channels are open, and the conductance at the Fermi energy takes on integer multiples of the conductance unit $G_{o}$. The conductance can attain high integer values that can be controlled by the width $N$ of the ZGNRs, in this way defining a multi-channel quantum wire ``current filter". Therefore, the system has potential to be employed in technological applications.

In the second case, for even $N$, the ZGNRs exhibit an insulator phase, with the gap at the Fermi energy decreasing logarithmically with an increase in $N$ and tending asymptotically to zero, leading to the ZGNR developing an insulator-metal transition. This is a kind of Lifshitz-type transition, but here the topology of the Fermi surface changes in a discrete way. As discussed earlier, we did not identify any real system in which the N3 and NN hopping have opposite signs. However, we think that one possibility for its experimental realization is employing optical lattices of ultracold atoms, as discussed in a recent paper \cite{Zhang2015}, where the authors described an experimental realization of tunable optical sawtooth and zigzag lattices, with precise control over the intra- and inter-unit-cell hopping. For odd $N$, the conductance exhibits a behavior similar to metallic armchair nanoribbons.

We also calculated the conductance of a ZGNR two-terminal device as a function of the hopping $t_{c}$ that connects the central cell and the leads, as indicated in Fig. \ref{Device12}b. We showed that the system can be tuned to an insulator-metal transition, with the conductance at the Fermi energy exhibiting a staircase behavior as a function of 
$t_{c}$ and assuming all the possible integer multiples of the conductance unit ($G_{o}$), compatible with the width of the ZGNRs.

Considering only a NN tight-binding calculation, the ZGNRs edge states were theoretically predicted \cite{Nakada96} to couple ferromagnetically along their edges and antiferromagnetically between them. In this case, the Stoner criterion is always satisfied, due to the moderate electronic correlation $U \simeq 0.8t$ \cite{Jung09} present in those ZGNRs, and the strong peak at the Fermi energy. However, when N3 hopping is included in the Hamiltonian, the peak at the Fermi energy is flattened, as indicated in 
Figs. (\ref{DFT_band_par},\ref{band_dos_cond_1})b, and the Stoner criterion is not satisfied for ZGNRs with $N=(2,3)$. Therefore, the magnetic order could not be developed at their edges. These conclusions are also confirmed by  LSDA simulations, but  LSDA is not able to resolve the fundamental state of these ZGNRs into NM, F, and AF configurations. One possibility to go further, is to employ the GW approximation,  that includes a screened-exchange self energy or a more sophisticated method that includes  local quantum fluctuations,  like DMFT combined with GW approximation  \cite{Tomczak2017}. Those methods could contribute to elucidate the real magnetic nature of these narrow zigzag nanoribbons.

The possibility of the existence of ZGNRs without magnetic order constitutes a new problem in the area and, after the recent synthesis of $6$-ZGNR \cite{Pascal16}, our work could stimulate experimental groups to improve the techniques to synthesize these ZGNRs and measure their spin-polarized edge states.

\section{Acknowledgments}

We thank CAPES and CNPq for the support of this
work. J.H.Correa and M.S.Figueira also thanks Profs. G. B. Martins, M. Guassi
and G. Diniz for helpful discussions. Part of the calculations were performed at the HPC/UFABC and  HPC/IFUSP.

\bibliography{biblio_PLA}

\begin{thebibliography}{48}%
\makeatletter
\providecommand \@ifxundefined [1]{%
 \@ifx{#1\undefined}
}%
\providecommand \@ifnum [1]{%
 \ifnum #1\expandafter \@firstoftwo
 \else \expandafter \@secondoftwo
 \fi
}%
\providecommand \@ifx [1]{%
 \ifx #1\expandafter \@firstoftwo
 \else \expandafter \@secondoftwo
 \fi
}%
\providecommand \natexlab [1]{#1}%
\providecommand \enquote  [1]{``#1''}%
\providecommand \bibnamefont  [1]{#1}%
\providecommand \bibfnamefont [1]{#1}%
\providecommand \citenamefont [1]{#1}%
\providecommand \href@noop [0]{\@secondoftwo}%
\providecommand \href [0]{\begingroup \@sanitize@url \@href}%
\providecommand \@href[1]{\@@startlink{#1}\@@href}%
\providecommand \@@href[1]{\endgroup#1\@@endlink}%
\providecommand \@sanitize@url [0]{\catcode `\\12\catcode `\$12\catcode
  `\&12\catcode `\#12\catcode `\^12\catcode `\_12\catcode `\%12\relax}%
\providecommand \@@startlink[1]{}%
\providecommand \@@endlink[0]{}%
\providecommand \url  [0]{\begingroup\@sanitize@url \@url }%
\providecommand \@url [1]{\endgroup\@href {#1}{\urlprefix }}%
\providecommand \urlprefix  [0]{URL }%
\providecommand \Eprint [0]{\href }%
\providecommand \doibase [0]{http://dx.doi.org/}%
\providecommand \selectlanguage [0]{\@gobble}%
\providecommand \bibinfo  [0]{\@secondoftwo}%
\providecommand \bibfield  [0]{\@secondoftwo}%
\providecommand \translation [1]{[#1]}%
\providecommand \BibitemOpen [0]{}%
\providecommand \bibitemStop [0]{}%
\providecommand \bibitemNoStop [0]{.\EOS\space}%
\providecommand \EOS [0]{\spacefactor3000\relax}%
\providecommand \BibitemShut  [1]{\csname bibitem#1\endcsname}%
\let\auto@bib@innerbib\@empty
%</preamble>
\bibitem [{\citenamefont {Novoselov}\ \emph {et~al.}(2004)\citenamefont
  {Novoselov}, \citenamefont {Geim}, \citenamefont {Morozov}, \citenamefont
  {Jiang}, \citenamefont {Zhang}, \citenamefont {Dubonos}, \citenamefont
  {Grigorieva},\ and\ \citenamefont {Firsov}}]{Novoselov666}%
  \BibitemOpen
  \bibfield  {author} {\bibinfo {author} {\bibfnamefont {K.~S.}\ \bibnamefont
  {Novoselov}}, \bibinfo {author} {\bibfnamefont {A.~K.}\ \bibnamefont {Geim}},
  \bibinfo {author} {\bibfnamefont {S.~V.}\ \bibnamefont {Morozov}}, \bibinfo
  {author} {\bibfnamefont {D.}~\bibnamefont {Jiang}}, \bibinfo {author}
  {\bibfnamefont {Y.}~\bibnamefont {Zhang}}, \bibinfo {author} {\bibfnamefont
  {S.~V.}\ \bibnamefont {Dubonos}}, \bibinfo {author} {\bibfnamefont {I.~V.}\
  \bibnamefont {Grigorieva}}, \ and\ \bibinfo {author} {\bibfnamefont {A.~A.}\
  \bibnamefont {Firsov}},\ }\href {\doibase 10.1126/science.1102896} {\bibfield
   {journal} {\bibinfo  {journal} {Science}\ }\textbf {\bibinfo {volume}
  {306}},\ \bibinfo {pages} {666} (\bibinfo {year} {2004})}\BibitemShut
  {NoStop}%
\bibitem [{\citenamefont {Castro~Neto}\ \emph {et~al.}(2009)\citenamefont
  {Castro~Neto}, \citenamefont {Guinea}, \citenamefont {Peres}, \citenamefont
  {Novoselov},\ and\ \citenamefont {Geim}}]{CastroNeto09}%
  \BibitemOpen
  \bibfield  {author} {\bibinfo {author} {\bibfnamefont {A.~H.}\ \bibnamefont
  {Castro~Neto}}, \bibinfo {author} {\bibfnamefont {F.}~\bibnamefont {Guinea}},
  \bibinfo {author} {\bibfnamefont {N.~M.~R.}\ \bibnamefont {Peres}}, \bibinfo
  {author} {\bibfnamefont {K.~S.}\ \bibnamefont {Novoselov}}, \ and\ \bibinfo
  {author} {\bibfnamefont {A.~K.}\ \bibnamefont {Geim}},\ }\href {\doibase
  10.1103/RevModPhys.81.109} {\bibfield  {journal} {\bibinfo  {journal} {Rev.
  Mod. Phys.}\ }\textbf {\bibinfo {volume} {81}},\ \bibinfo {pages} {109}
  (\bibinfo {year} {2009})}\BibitemShut {NoStop}%
\bibitem [{\citenamefont {Gunlycke}\ \emph {et~al.}(2007)\citenamefont
  {Gunlycke}, \citenamefont {Areshkin}, \citenamefont {Li}, \citenamefont
  {Mintmire},\ and\ \citenamefont {White}}]{Gunlycke207}%
  \BibitemOpen
  \bibfield  {author} {\bibinfo {author} {\bibfnamefont {D.}~\bibnamefont
  {Gunlycke}}, \bibinfo {author} {\bibfnamefont {D.~A.}\ \bibnamefont
  {Areshkin}}, \bibinfo {author} {\bibfnamefont {J.}~\bibnamefont {Li}},
  \bibinfo {author} {\bibfnamefont {J.~W.}\ \bibnamefont {Mintmire}}, \ and\
  \bibinfo {author} {\bibfnamefont {C.~T.}\ \bibnamefont {White}},\ }\href
  {\doibase 10.1021/nl0717917} {\bibfield  {journal} {\bibinfo  {journal} {Nano
  Letters}\ }\textbf {\bibinfo {volume} {7}},\ \bibinfo {pages} {3608}
  (\bibinfo {year} {2007})},\ \bibinfo {note} {pMID: 18004900},\ \Eprint
  {http://arxiv.org/abs/https://doi.org/10.1021/nl0717917}
  {https://doi.org/10.1021/nl0717917} \BibitemShut {NoStop}%
\bibitem [{\citenamefont {Areshkin}\ \emph {et~al.}(2007)\citenamefont
  {Areshkin}, \citenamefont {Gunlycke},\ and\ \citenamefont
  {White}}]{Gunlycke07}%
  \BibitemOpen
  \bibfield  {author} {\bibinfo {author} {\bibfnamefont {D.~A.}\ \bibnamefont
  {Areshkin}}, \bibinfo {author} {\bibfnamefont {D.}~\bibnamefont {Gunlycke}},
  \ and\ \bibinfo {author} {\bibfnamefont {C.~T.}\ \bibnamefont {White}},\
  }\href {\doibase 10.1021/nl062132h} {\bibfield  {journal} {\bibinfo
  {journal} {Nano Letters}\ }\textbf {\bibinfo {volume} {7}},\ \bibinfo {pages}
  {204} (\bibinfo {year} {2007})},\ \bibinfo {note} {pMID: 17212465},\ \Eprint
  {http://arxiv.org/abs/https://doi.org/10.1021/nl062132h}
  {https://doi.org/10.1021/nl062132h} \BibitemShut {NoStop}%
\bibitem [{\citenamefont {Celis}\ \emph {et~al.}(2016)\citenamefont {Celis},
  \citenamefont {Nair}, \citenamefont {Taleb-Ibrahimi}, \citenamefont {Conrad},
  \citenamefont {Berger}, \citenamefont {de~Heer},\ and\ \citenamefont
  {Tejeda}}]{Celis2016}%
  \BibitemOpen
  \bibfield  {author} {\bibinfo {author} {\bibfnamefont {A.}~\bibnamefont
  {Celis}}, \bibinfo {author} {\bibfnamefont {M.~N.}\ \bibnamefont {Nair}},
  \bibinfo {author} {\bibfnamefont {A.}~\bibnamefont {Taleb-Ibrahimi}},
  \bibinfo {author} {\bibfnamefont {E.~H.}\ \bibnamefont {Conrad}}, \bibinfo
  {author} {\bibfnamefont {C.}~\bibnamefont {Berger}}, \bibinfo {author}
  {\bibfnamefont {W.~A.}\ \bibnamefont {de~Heer}}, \ and\ \bibinfo {author}
  {\bibfnamefont {A.}~\bibnamefont {Tejeda}},\ }\href
  {http://stacks.iop.org/0022-3727/49/i=14/a=143001} {\bibfield  {journal}
  {\bibinfo  {journal} {Journal of Physics D: Applied Physics}\ }\textbf
  {\bibinfo {volume} {49}},\ \bibinfo {pages} {143001} (\bibinfo {year}
  {2016})}\BibitemShut {NoStop}%
\bibitem [{\citenamefont {Ruffieux}\ \emph {et~al.}(2016)\citenamefont
  {Ruffieux}, \citenamefont {Wang}, \citenamefont {Yang}, \citenamefont
  {S\'anchez-S\'anchez}, \citenamefont {Liu}, \citenamefont {Dienel},
  \citenamefont {Talirz}, \citenamefont {Shinde}, \citenamefont {Pignedoli},
  \citenamefont {Passerone}, \citenamefont {Dumslaff}, \citenamefont {Feng},
  \citenamefont {M\"ullen},\ and\ \citenamefont {Fasel}}]{Pascal16}%
  \BibitemOpen
  \bibfield  {author} {\bibinfo {author} {\bibfnamefont {P.}~\bibnamefont
  {Ruffieux}}, \bibinfo {author} {\bibfnamefont {S.}~\bibnamefont {Wang}},
  \bibinfo {author} {\bibfnamefont {B.}~\bibnamefont {Yang}}, \bibinfo {author}
  {\bibfnamefont {C.}~\bibnamefont {S\'anchez-S\'anchez}}, \bibinfo {author}
  {\bibfnamefont {J.}~\bibnamefont {Liu}}, \bibinfo {author} {\bibfnamefont
  {T.}~\bibnamefont {Dienel}}, \bibinfo {author} {\bibfnamefont
  {L.}~\bibnamefont {Talirz}}, \bibinfo {author} {\bibfnamefont
  {P.}~\bibnamefont {Shinde}}, \bibinfo {author} {\bibfnamefont {C.~A.}\
  \bibnamefont {Pignedoli}}, \bibinfo {author} {\bibfnamefont {D.}~\bibnamefont
  {Passerone}}, \bibinfo {author} {\bibfnamefont {T.}~\bibnamefont {Dumslaff}},
  \bibinfo {author} {\bibfnamefont {X.}~\bibnamefont {Feng}}, \bibinfo {author}
  {\bibfnamefont {K.}~\bibnamefont {M\"ullen}}, \ and\ \bibinfo {author}
  {\bibfnamefont {R.}~\bibnamefont {Fasel}},\ }\href
  {http://dx.doi.org/10.1038/nature17151} {\bibfield  {journal} {\bibinfo
  {journal} {Nature}\ }\textbf {\bibinfo {volume} {531}},\ \bibinfo {pages}
  {489} (\bibinfo {year} {2016})}\BibitemShut {NoStop}%
\bibitem [{\citenamefont {Nakada}\ \emph {et~al.}(1996)\citenamefont {Nakada},
  \citenamefont {Fujita}, \citenamefont {Dresselhaus},\ and\ \citenamefont
  {Dresselhaus}}]{Nakada96}%
  \BibitemOpen
  \bibfield  {author} {\bibinfo {author} {\bibfnamefont {K.}~\bibnamefont
  {Nakada}}, \bibinfo {author} {\bibfnamefont {M.}~\bibnamefont {Fujita}},
  \bibinfo {author} {\bibfnamefont {G.}~\bibnamefont {Dresselhaus}}, \ and\
  \bibinfo {author} {\bibfnamefont {M.~S.}\ \bibnamefont {Dresselhaus}},\
  }\href {\doibase 10.1103/PhysRevB.54.17954} {\bibfield  {journal} {\bibinfo
  {journal} {Phys. Rev. B}\ }\textbf {\bibinfo {volume} {54}},\ \bibinfo
  {pages} {17954} (\bibinfo {year} {1996})}\BibitemShut {NoStop}%
\bibitem [{\citenamefont {Kivelson}\ and\ \citenamefont
  {Chapman}(1983)}]{Kivelson1983}%
  \BibitemOpen
  \bibfield  {author} {\bibinfo {author} {\bibfnamefont {S.}~\bibnamefont
  {Kivelson}}\ and\ \bibinfo {author} {\bibfnamefont {O.~L.}\ \bibnamefont
  {Chapman}},\ }\href {\doibase 10.1103/PhysRevB.28.7236} {\bibfield  {journal}
  {\bibinfo  {journal} {Phys. Rev. B}\ }\textbf {\bibinfo {volume} {28}},\
  \bibinfo {pages} {7236} (\bibinfo {year} {1983})}\BibitemShut {NoStop}%
\bibitem [{\citenamefont {Karakonstantakis}\ \emph {et~al.}(2013)\citenamefont
  {Karakonstantakis}, \citenamefont {Liu}, \citenamefont {Thomale},\ and\
  \citenamefont {Kivelson}}]{Kivelson2013}%
  \BibitemOpen
  \bibfield  {author} {\bibinfo {author} {\bibfnamefont {G.}~\bibnamefont
  {Karakonstantakis}}, \bibinfo {author} {\bibfnamefont {L.}~\bibnamefont
  {Liu}}, \bibinfo {author} {\bibfnamefont {R.}~\bibnamefont {Thomale}}, \ and\
  \bibinfo {author} {\bibfnamefont {S.~A.}\ \bibnamefont {Kivelson}},\ }\href
  {\doibase 10.1103/PhysRevB.88.224512} {\bibfield  {journal} {\bibinfo
  {journal} {Phys. Rev. B}\ }\textbf {\bibinfo {volume} {88}},\ \bibinfo
  {pages} {224512} (\bibinfo {year} {2013})}\BibitemShut {NoStop}%
\bibitem [{\citenamefont {Schmitteckert}\ \emph {et~al.}(2017)\citenamefont
  {Schmitteckert}, \citenamefont {Thomale}, \citenamefont {Koryt\'ar},\ and\
  \citenamefont {Evers}}]{Korytar17}%
  \BibitemOpen
  \bibfield  {author} {\bibinfo {author} {\bibfnamefont {P.}~\bibnamefont
  {Schmitteckert}}, \bibinfo {author} {\bibfnamefont {R.}~\bibnamefont
  {Thomale}}, \bibinfo {author} {\bibfnamefont {R.}~\bibnamefont {Koryt\'ar}},
  \ and\ \bibinfo {author} {\bibfnamefont {F.}~\bibnamefont {Evers}},\ }\href
  {\doibase 10.1063/1.4975319} {\bibfield  {journal} {\bibinfo  {journal} {The
  Journal of Chemical Physics}\ }\textbf {\bibinfo {volume} {146}},\ \bibinfo
  {pages} {092320} (\bibinfo {year} {2017})}\BibitemShut {NoStop}%
\bibitem [{\citenamefont {van Ostaay}\ \emph {et~al.}(2011)\citenamefont {van
  Ostaay}, \citenamefont {Akhmerov}, \citenamefont {Beenakker},\ and\
  \citenamefont {Wimmer}}]{Ostaay11}%
  \BibitemOpen
  \bibfield  {author} {\bibinfo {author} {\bibfnamefont {J.~A.~M.}\
  \bibnamefont {van Ostaay}}, \bibinfo {author} {\bibfnamefont {A.~R.}\
  \bibnamefont {Akhmerov}}, \bibinfo {author} {\bibfnamefont {C.~W.~J.}\
  \bibnamefont {Beenakker}}, \ and\ \bibinfo {author} {\bibfnamefont
  {M.}~\bibnamefont {Wimmer}},\ }\href {\doibase 10.1103/PhysRevB.84.195434}
  {\bibfield  {journal} {\bibinfo  {journal} {Phys. Rev. B}\ }\textbf {\bibinfo
  {volume} {84}},\ \bibinfo {pages} {195434} (\bibinfo {year}
  {2011})}\BibitemShut {NoStop}%
\bibitem [{\citenamefont {Ziman}(1972)}]{Ziman72}%
  \BibitemOpen
  \bibfield  {author} {\bibinfo {author} {\bibfnamefont {J.}~\bibnamefont
  {Ziman}},\ }\href@noop {} {\emph {\bibinfo {title} {{Principles of the Theory
  of Solids - Second edition}}}}\ (\bibinfo  {publisher} {Cambridge University
  Press},\ \bibinfo {address} {London, England},\ \bibinfo {year}
  {1972})\BibitemShut {NoStop}%
\bibitem [{\citenamefont {Teodorescu}\ and\ \citenamefont
  {Lungu}(2008)}]{Teodorescu008}%
  \BibitemOpen
  \bibfield  {author} {\bibinfo {author} {\bibfnamefont {C.~M.}\ \bibnamefont
  {Teodorescu}}\ and\ \bibinfo {author} {\bibfnamefont {G.~A.}\ \bibnamefont
  {Lungu}},\ }\href
  {https://joam.inoe.ro/index.php?option=magazine&op=view&idu=1752&catid=32}
  {\bibfield  {journal} {\bibinfo  {journal} {Journal of optoelectronics and
  advanced materials}\ }\textbf {\bibinfo {volume} {10}},\ \bibinfo {pages}
  {3058} (\bibinfo {year} {2008})}\BibitemShut {NoStop}%
\bibitem [{\citenamefont {Bena}\ and\ \citenamefont
  {Simon}(2011)}]{Cristina2011}%
  \BibitemOpen
  \bibfield  {author} {\bibinfo {author} {\bibfnamefont {C.}~\bibnamefont
  {Bena}}\ and\ \bibinfo {author} {\bibfnamefont {L.}~\bibnamefont {Simon}},\
  }\href {\doibase 10.1103/PhysRevB.83.115404} {\bibfield  {journal} {\bibinfo
  {journal} {Phys. Rev. B}\ }\textbf {\bibinfo {volume} {83}},\ \bibinfo
  {pages} {115404} (\bibinfo {year} {2011})}\BibitemShut {NoStop}%
\bibitem [{\citenamefont {Brey}(2015)}]{Brey15}%
  \BibitemOpen
  \bibfield  {author} {\bibinfo {author} {\bibfnamefont {L.}~\bibnamefont
  {Brey}},\ }\href {\doibase 10.1103/PhysRevB.92.235444} {\bibfield  {journal}
  {\bibinfo  {journal} {Phys. Rev. B}\ }\textbf {\bibinfo {volume} {92}},\
  \bibinfo {pages} {235444} (\bibinfo {year} {2015})}\BibitemShut {NoStop}%
\bibitem [{\citenamefont {Carvalho}\ \emph {et~al.}(2014)\citenamefont
  {Carvalho}, \citenamefont {Warnes},\ and\ \citenamefont
  {Lewenkopf}}]{Caio14}%
  \BibitemOpen
  \bibfield  {author} {\bibinfo {author} {\bibfnamefont {A.~R.}\ \bibnamefont
  {Carvalho}}, \bibinfo {author} {\bibfnamefont {J.~H.}\ \bibnamefont
  {Warnes}}, \ and\ \bibinfo {author} {\bibfnamefont {C.~H.}\ \bibnamefont
  {Lewenkopf}},\ }\href {\doibase 10.1103/PhysRevB.89.245444} {\bibfield
  {journal} {\bibinfo  {journal} {Phys. Rev. B}\ }\textbf {\bibinfo {volume}
  {89}},\ \bibinfo {pages} {245444} (\bibinfo {year} {2014})}\BibitemShut
  {NoStop}%
\bibitem [{\citenamefont {Kane}\ and\ \citenamefont
  {Mele}(2005)}]{Kane_Mele_05}%
  \BibitemOpen
  \bibfield  {author} {\bibinfo {author} {\bibfnamefont {C.~L.}\ \bibnamefont
  {Kane}}\ and\ \bibinfo {author} {\bibfnamefont {E.~J.}\ \bibnamefont
  {Mele}},\ }\href {\doibase 10.1103/PhysRevLett.95.226801} {\bibfield
  {journal} {\bibinfo  {journal} {Phys. Rev. Lett.}\ }\textbf {\bibinfo
  {volume} {95}},\ \bibinfo {pages} {226801} (\bibinfo {year}
  {2005})}\BibitemShut {NoStop}%
\bibitem [{\citenamefont {Reich}\ \emph {et~al.}(2002)\citenamefont {Reich},
  \citenamefont {Maultzsch}, \citenamefont {Thomsen},\ and\ \citenamefont
  {Ordej\'on}}]{Reich2002}%
  \BibitemOpen
  \bibfield  {author} {\bibinfo {author} {\bibfnamefont {S.}~\bibnamefont
  {Reich}}, \bibinfo {author} {\bibfnamefont {J.}~\bibnamefont {Maultzsch}},
  \bibinfo {author} {\bibfnamefont {C.}~\bibnamefont {Thomsen}}, \ and\
  \bibinfo {author} {\bibfnamefont {P.}~\bibnamefont {Ordej\'on}},\ }\href
  {\doibase 10.1103/PhysRevB.66.035412} {\bibfield  {journal} {\bibinfo
  {journal} {Phys. Rev. B}\ }\textbf {\bibinfo {volume} {66}},\ \bibinfo
  {pages} {035412} (\bibinfo {year} {2002})}\BibitemShut {NoStop}%
\bibitem [{\citenamefont {Gunlycke}\ and\ \citenamefont
  {White}(2008)}]{Gunlycke08}%
  \BibitemOpen
  \bibfield  {author} {\bibinfo {author} {\bibfnamefont {D.}~\bibnamefont
  {Gunlycke}}\ and\ \bibinfo {author} {\bibfnamefont {C.~T.}\ \bibnamefont
  {White}},\ }\href {\doibase 10.1103/PhysRevB.77.115116} {\bibfield  {journal}
  {\bibinfo  {journal} {Phys. Rev. B}\ }\textbf {\bibinfo {volume} {77}},\
  \bibinfo {pages} {115116} (\bibinfo {year} {2008})}\BibitemShut {NoStop}%
\bibitem [{\citenamefont {Hancock}\ \emph {et~al.}(2010)\citenamefont
  {Hancock}, \citenamefont {Uppstu}, \citenamefont {Saloriutta}, \citenamefont
  {Harju},\ and\ \citenamefont {Puska}}]{Hancock10}%
  \BibitemOpen
  \bibfield  {author} {\bibinfo {author} {\bibfnamefont {Y.}~\bibnamefont
  {Hancock}}, \bibinfo {author} {\bibfnamefont {A.}~\bibnamefont {Uppstu}},
  \bibinfo {author} {\bibfnamefont {K.}~\bibnamefont {Saloriutta}}, \bibinfo
  {author} {\bibfnamefont {A.}~\bibnamefont {Harju}}, \ and\ \bibinfo {author}
  {\bibfnamefont {M.~J.}\ \bibnamefont {Puska}},\ }\href {\doibase
  10.1103/PhysRevB.81.245402} {\bibfield  {journal} {\bibinfo  {journal} {Phys.
  Rev. B}\ }\textbf {\bibinfo {volume} {81}},\ \bibinfo {pages} {245402}
  (\bibinfo {year} {2010})}\BibitemShut {NoStop}%
\bibitem [{\citenamefont {Wu}\ and\ \citenamefont {Childs}(2010)}]{Wu2010}%
  \BibitemOpen
  \bibfield  {author} {\bibinfo {author} {\bibfnamefont {Y.}~\bibnamefont
  {Wu}}\ and\ \bibinfo {author} {\bibfnamefont {P.}~\bibnamefont {Childs}},\
  }\href {\doibase 10.1007/s11671-010-9791-y} {\bibfield  {journal} {\bibinfo
  {journal} {Nanoscale Res Lett}\ }\textbf {\bibinfo {volume} {6}},\ \bibinfo
  {pages} {62} (\bibinfo {year} {2010})}\BibitemShut {NoStop}%
\bibitem [{\citenamefont {Kundu}(2011)}]{Kundu11}%
  \BibitemOpen
  \bibfield  {author} {\bibinfo {author} {\bibfnamefont {R.}~\bibnamefont
  {Kundu}},\ }\href {\doibase 10.1142/S0217984911025663} {\bibfield  {journal}
  {\bibinfo  {journal} {Modern Physics Letters B}\ }\textbf {\bibinfo {volume}
  {25}},\ \bibinfo {pages} {163} (\bibinfo {year} {2011})}\BibitemShut
  {NoStop}%
\bibitem [{\citenamefont {Tran}\ \emph {et~al.}(2017)\citenamefont {Tran},
  \citenamefont {Saint-Martin}, \citenamefont {Dollfus},\ and\ \citenamefont
  {Volz}}]{Tran17}%
  \BibitemOpen
  \bibfield  {author} {\bibinfo {author} {\bibfnamefont {V.-T.}\ \bibnamefont
  {Tran}}, \bibinfo {author} {\bibfnamefont {J.}~\bibnamefont {Saint-Martin}},
  \bibinfo {author} {\bibfnamefont {P.}~\bibnamefont {Dollfus}}, \ and\
  \bibinfo {author} {\bibfnamefont {S.}~\bibnamefont {Volz}},\ }\href {\doibase
  10.1063/1.4994771} {\bibfield  {journal} {\bibinfo  {journal} {AIP Advances}\
  }\textbf {\bibinfo {volume} {7}},\ \bibinfo {pages} {075212} (\bibinfo {year}
  {2017})}\BibitemShut {NoStop}%
\bibitem [{\citenamefont {Hung}\ \emph {et~al.}(2013)\citenamefont {Hung},
  \citenamefont {Wang}, \citenamefont {Gu},\ and\ \citenamefont
  {Fiete}}]{Hung13}%
  \BibitemOpen
  \bibfield  {author} {\bibinfo {author} {\bibfnamefont {H.-H.}\ \bibnamefont
  {Hung}}, \bibinfo {author} {\bibfnamefont {L.}~\bibnamefont {Wang}}, \bibinfo
  {author} {\bibfnamefont {Z.-C.}\ \bibnamefont {Gu}}, \ and\ \bibinfo {author}
  {\bibfnamefont {G.~A.}\ \bibnamefont {Fiete}},\ }\href {\doibase
  10.1103/PhysRevB.87.121113} {\bibfield  {journal} {\bibinfo  {journal} {Phys.
  Rev. B}\ }\textbf {\bibinfo {volume} {87}},\ \bibinfo {pages} {121113}
  (\bibinfo {year} {2013})}\BibitemShut {NoStop}%
\bibitem [{\citenamefont {Hung}\ \emph {et~al.}(2014)\citenamefont {Hung},
  \citenamefont {Chua}, \citenamefont {Wang},\ and\ \citenamefont
  {Fiete}}]{Hung14}%
  \BibitemOpen
  \bibfield  {author} {\bibinfo {author} {\bibfnamefont {H.-H.}\ \bibnamefont
  {Hung}}, \bibinfo {author} {\bibfnamefont {V.}~\bibnamefont {Chua}}, \bibinfo
  {author} {\bibfnamefont {L.}~\bibnamefont {Wang}}, \ and\ \bibinfo {author}
  {\bibfnamefont {G.~A.}\ \bibnamefont {Fiete}},\ }\href {\doibase
  10.1103/PhysRevB.89.235104} {\bibfield  {journal} {\bibinfo  {journal} {Phys.
  Rev. B}\ }\textbf {\bibinfo {volume} {89}},\ \bibinfo {pages} {235104}
  (\bibinfo {year} {2014})}\BibitemShut {NoStop}%
\bibitem [{\citenamefont {Chen}\ \emph {et~al.}(2015)\citenamefont {Chen},
  \citenamefont {Hung}, \citenamefont {Su}, \citenamefont {Fiete},\ and\
  \citenamefont {Ting}}]{Chen15}%
  \BibitemOpen
  \bibfield  {author} {\bibinfo {author} {\bibfnamefont {Y.-H.}\ \bibnamefont
  {Chen}}, \bibinfo {author} {\bibfnamefont {H.-H.}\ \bibnamefont {Hung}},
  \bibinfo {author} {\bibfnamefont {G.}~\bibnamefont {Su}}, \bibinfo {author}
  {\bibfnamefont {G.~A.}\ \bibnamefont {Fiete}}, \ and\ \bibinfo {author}
  {\bibfnamefont {C.~S.}\ \bibnamefont {Ting}},\ }\href {\doibase
  10.1103/PhysRevB.91.045122} {\bibfield  {journal} {\bibinfo  {journal} {Phys.
  Rev. B}\ }\textbf {\bibinfo {volume} {91}},\ \bibinfo {pages} {045122}
  (\bibinfo {year} {2015})}\BibitemShut {NoStop}%
\bibitem [{\citenamefont {Yazyev}(2010)}]{Yazyev2010}%
  \BibitemOpen
  \bibfield  {author} {\bibinfo {author} {\bibfnamefont {O.~V.}\ \bibnamefont
  {Yazyev}},\ }\href {http://stacks.iop.org/0034-4885/73/i=5/a=056501}
  {\bibfield  {journal} {\bibinfo  {journal} {Reports on Progress in Physics}\
  }\textbf {\bibinfo {volume} {73}},\ \bibinfo {pages} {056501} (\bibinfo
  {year} {2010})}\BibitemShut {NoStop}%
\bibitem [{\citenamefont {Guassi}\ \emph {et~al.}(2015)\citenamefont {Guassi},
  \citenamefont {Diniz}, \citenamefont {Sandler},\ and\ \citenamefont
  {Qu}}]{Marcos2015}%
  \BibitemOpen
  \bibfield  {author} {\bibinfo {author} {\bibfnamefont {M.~R.}\ \bibnamefont
  {Guassi}}, \bibinfo {author} {\bibfnamefont {G.~S.}\ \bibnamefont {Diniz}},
  \bibinfo {author} {\bibfnamefont {N.}~\bibnamefont {Sandler}}, \ and\
  \bibinfo {author} {\bibfnamefont {F.}~\bibnamefont {Qu}},\ }\href {\doibase
  10.1103/PhysRevB.92.075426} {\bibfield  {journal} {\bibinfo  {journal} {Phys.
  Rev. B}\ }\textbf {\bibinfo {volume} {92}},\ \bibinfo {pages} {075426}
  (\bibinfo {year} {2015})}\BibitemShut {NoStop}%
\bibitem [{\citenamefont {Sancho}\ \emph {et~al.}(1984)\citenamefont {Sancho},
  \citenamefont {Sancho},\ and\ \citenamefont {Rubio}}]{Rubio}%
  \BibitemOpen
  \bibfield  {author} {\bibinfo {author} {\bibfnamefont {M.~P.~L.}\
  \bibnamefont {Sancho}}, \bibinfo {author} {\bibfnamefont {J.~M.~L.}\
  \bibnamefont {Sancho}}, \ and\ \bibinfo {author} {\bibfnamefont
  {J.}~\bibnamefont {Rubio}},\ }\href
  {http://stacks.iop.org/0305-4608/14/i=5/a=016} {\bibfield  {journal}
  {\bibinfo  {journal} {Journal of Physics F: Metal Physics}\ }\textbf
  {\bibinfo {volume} {14}},\ \bibinfo {pages} {1205} (\bibinfo {year}
  {1984})}\BibitemShut {NoStop}%
\bibitem [{\citenamefont {Diniz}\ \emph {et~al.}(2014)\citenamefont {Diniz},
  \citenamefont {Guassi},\ and\ \citenamefont {Qu}}]{ginetom}%
  \BibitemOpen
  \bibfield  {author} {\bibinfo {author} {\bibfnamefont {G.~S.}\ \bibnamefont
  {Diniz}}, \bibinfo {author} {\bibfnamefont {M.~R.}\ \bibnamefont {Guassi}}, \
  and\ \bibinfo {author} {\bibfnamefont {F.}~\bibnamefont {Qu}},\ }\href
  {\doibase 10.1063/1.4896251} {\bibfield  {journal} {\bibinfo  {journal}
  {Journal of Applied Physics}\ }\textbf {\bibinfo {volume} {116}},\ \bibinfo
  {pages} {113705} (\bibinfo {year} {2014})}\BibitemShut {NoStop}%
\bibitem [{\citenamefont {Nardelli}(1999)}]{nardelli}%
  \BibitemOpen
  \bibfield  {author} {\bibinfo {author} {\bibfnamefont {M.~B.}\ \bibnamefont
  {Nardelli}},\ }\href {\doibase 10.1103/PhysRevB.60.7828} {\bibfield
  {journal} {\bibinfo  {journal} {Phys. Rev. B}\ }\textbf {\bibinfo {volume}
  {60}},\ \bibinfo {pages} {7828} (\bibinfo {year} {1999})}\BibitemShut
  {NoStop}%
\bibitem [{\citenamefont {Soler}\ \emph {et~al.}(2002)\citenamefont {Soler},
  \citenamefont {Artacho}, \citenamefont {Gale}, \citenamefont {Garc\'{i}a},
  \citenamefont {Junquera}, \citenamefont {Ordej\'{o}n},\ and\ \citenamefont
  {S\'{a}nchez-Portal}}]{SIESTA}%
  \BibitemOpen
  \bibfield  {author} {\bibinfo {author} {\bibfnamefont {J.~M.}\ \bibnamefont
  {Soler}}, \bibinfo {author} {\bibfnamefont {E.}~\bibnamefont {Artacho}},
  \bibinfo {author} {\bibfnamefont {J.~D.}\ \bibnamefont {Gale}}, \bibinfo
  {author} {\bibfnamefont {A.}~\bibnamefont {Garc\'{i}a}}, \bibinfo {author}
  {\bibfnamefont {J.}~\bibnamefont {Junquera}}, \bibinfo {author}
  {\bibfnamefont {P.}~\bibnamefont {Ordej\'{o}n}}, \ and\ \bibinfo {author}
  {\bibfnamefont {D.}~\bibnamefont {S\'{a}nchez-Portal}},\ }\href
  {http://stacks.iop.org/0953-8984/14/i=11/a=302} {\bibfield  {journal}
  {\bibinfo  {journal} {Journal of Physics: Condensed Matter}\ }\textbf
  {\bibinfo {volume} {14}},\ \bibinfo {pages} {2745} (\bibinfo {year}
  {2002})}\BibitemShut {NoStop}%
\bibitem [{\citenamefont {Koryt\'ar}\ \emph {et~al.}(2014)\citenamefont
  {Koryt\'ar}, \citenamefont {Xenioti}, \citenamefont {Schmitteckert},
  \citenamefont {Alouani},\ and\ \citenamefont {Evers}}]{Korytar2014}%
  \BibitemOpen
  \bibfield  {author} {\bibinfo {author} {\bibfnamefont {R.}~\bibnamefont
  {Koryt\'ar}}, \bibinfo {author} {\bibfnamefont {D.}~\bibnamefont {Xenioti}},
  \bibinfo {author} {\bibfnamefont {P.}~\bibnamefont {Schmitteckert}}, \bibinfo
  {author} {\bibfnamefont {M.}~\bibnamefont {Alouani}}, \ and\ \bibinfo
  {author} {\bibfnamefont {F.}~\bibnamefont {Evers}},\ }\href {\doibase
  doi:10.1038/ncomms6000} {\ \textbf {\bibinfo {volume} {5}},\ \bibinfo {pages}
  {5000} (\bibinfo {year} {2014})}\BibitemShut {NoStop}%
\bibitem [{\citenamefont {Liu}\ \emph {et~al.}(2011)\citenamefont {Liu},
  \citenamefont {Jiang},\ and\ \citenamefont {Yao}}]{Liu2011}%
  \BibitemOpen
  \bibfield  {author} {\bibinfo {author} {\bibfnamefont {C.-C.}\ \bibnamefont
  {Liu}}, \bibinfo {author} {\bibfnamefont {H.}~\bibnamefont {Jiang}}, \ and\
  \bibinfo {author} {\bibfnamefont {Y.}~\bibnamefont {Yao}},\ }\href {\doibase
  10.1103/PhysRevB.84.195430} {\bibfield  {journal} {\bibinfo  {journal} {Phys.
  Rev. B}\ }\textbf {\bibinfo {volume} {84}},\ \bibinfo {pages} {195430}
  (\bibinfo {year} {2011})}\BibitemShut {NoStop}%
\bibitem [{\citenamefont {Padova}\ \emph {et~al.}(2012)\citenamefont {Padova},
  \citenamefont {Perfetti}, \citenamefont {Olivieri}, \citenamefont
  {Quaresima}, \citenamefont {Ottaviani},\ and\ \citenamefont
  {Lay}}]{Paola_silicene}%
  \BibitemOpen
  \bibfield  {author} {\bibinfo {author} {\bibfnamefont {P.~D.}\ \bibnamefont
  {Padova}}, \bibinfo {author} {\bibfnamefont {P.}~\bibnamefont {Perfetti}},
  \bibinfo {author} {\bibfnamefont {B.}~\bibnamefont {Olivieri}}, \bibinfo
  {author} {\bibfnamefont {C.}~\bibnamefont {Quaresima}}, \bibinfo {author}
  {\bibfnamefont {C.}~\bibnamefont {Ottaviani}}, \ and\ \bibinfo {author}
  {\bibfnamefont {G.~L.}\ \bibnamefont {Lay}},\ }\href
  {http://stacks.iop.org/0953-8984/24/i=22/a=223001} {\bibfield  {journal}
  {\bibinfo  {journal} {Journal of Physics: Condensed Matter}\ }\textbf
  {\bibinfo {volume} {24}},\ \bibinfo {pages} {223001} (\bibinfo {year}
  {2012})}\BibitemShut {NoStop}%
\bibitem [{\citenamefont {Zhang}\ \emph {et~al.}(2016)\citenamefont {Zhang},
  \citenamefont {Bampoulis}, \citenamefont {Rudenko}, \citenamefont {Yao},
  \citenamefont {van Houselt}, \citenamefont {Poelsema}, \citenamefont
  {Katsnelson},\ and\ \citenamefont {Zandvliet}}]{Zhang2016}%
  \BibitemOpen
  \bibfield  {author} {\bibinfo {author} {\bibfnamefont {L.}~\bibnamefont
  {Zhang}}, \bibinfo {author} {\bibfnamefont {P.}~\bibnamefont {Bampoulis}},
  \bibinfo {author} {\bibfnamefont {A.~N.}\ \bibnamefont {Rudenko}}, \bibinfo
  {author} {\bibfnamefont {Q.}~\bibnamefont {Yao}}, \bibinfo {author}
  {\bibfnamefont {A.}~\bibnamefont {van Houselt}}, \bibinfo {author}
  {\bibfnamefont {B.}~\bibnamefont {Poelsema}}, \bibinfo {author}
  {\bibfnamefont {M.~I.}\ \bibnamefont {Katsnelson}}, \ and\ \bibinfo {author}
  {\bibfnamefont {H.~J.~W.}\ \bibnamefont {Zandvliet}},\ }\href {\doibase
  10.1103/PhysRevLett.116.256804} {\bibfield  {journal} {\bibinfo  {journal}
  {Phys. Rev. Lett.}\ }\textbf {\bibinfo {volume} {116}},\ \bibinfo {pages}
  {256804} (\bibinfo {year} {2016})}\BibitemShut {NoStop}%
\bibitem [{\citenamefont {Zhu}\ \emph {et~al.}(2015)\citenamefont {Zhu},
  \citenamefont {Chen}, \citenamefont {Xu}, \citenamefont {Gao}, \citenamefont
  {Guan}, \citenamefont {Liu}, \citenamefont {Qian}, \citenamefont {Zhang},\
  and\ \citenamefont {Jia}}]{Zhu2015}%
  \BibitemOpen
  \bibfield  {author} {\bibinfo {author} {\bibfnamefont {F.-f.}\ \bibnamefont
  {Zhu}}, \bibinfo {author} {\bibfnamefont {W.-j.}\ \bibnamefont {Chen}},
  \bibinfo {author} {\bibfnamefont {Y.}~\bibnamefont {Xu}}, \bibinfo {author}
  {\bibfnamefont {C.-l.}\ \bibnamefont {Gao}}, \bibinfo {author} {\bibfnamefont
  {D.-d.}\ \bibnamefont {Guan}}, \bibinfo {author} {\bibfnamefont {C.-h.}\
  \bibnamefont {Liu}}, \bibinfo {author} {\bibfnamefont {D.}~\bibnamefont
  {Qian}}, \bibinfo {author} {\bibfnamefont {S.-C.}\ \bibnamefont {Zhang}}, \
  and\ \bibinfo {author} {\bibfnamefont {J.-f.}\ \bibnamefont {Jia}},\ }\href
  {http://dx.doi.org/10.1038/nmat4384} {\bibfield  {journal} {\bibinfo
  {journal} {Nat Mater}\ }\textbf {\bibinfo {volume} {14}},\ \bibinfo {pages}
  {1020} (\bibinfo {year} {2015})}\BibitemShut {NoStop}%
\bibitem [{\citenamefont {Allerdt}\ \emph {et~al.}(2017)\citenamefont
  {Allerdt}, \citenamefont {Feiguin},\ and\ \citenamefont
  {Martins}}]{Martins17}%
  \BibitemOpen
  \bibfield  {author} {\bibinfo {author} {\bibfnamefont {A.}~\bibnamefont
  {Allerdt}}, \bibinfo {author} {\bibfnamefont {A.~E.}\ \bibnamefont
  {Feiguin}}, \ and\ \bibinfo {author} {\bibfnamefont {G.~B.}\ \bibnamefont
  {Martins}},\ }\href {\doibase 10.1103/PhysRevB.96.035109} {\bibfield
  {journal} {\bibinfo  {journal} {Phys. Rev. B}\ }\textbf {\bibinfo {volume}
  {96}},\ \bibinfo {pages} {035109} (\bibinfo {year} {2017})}\BibitemShut
  {NoStop}%
\bibitem [{\citenamefont {Jung}\ and\ \citenamefont
  {MacDonald}(2009)}]{Jung09}%
  \BibitemOpen
  \bibfield  {author} {\bibinfo {author} {\bibfnamefont {J.}~\bibnamefont
  {Jung}}\ and\ \bibinfo {author} {\bibfnamefont {A.~H.}\ \bibnamefont
  {MacDonald}},\ }\href {\doibase 10.1103/PhysRevB.79.235433} {\bibfield
  {journal} {\bibinfo  {journal} {Phys. Rev. B}\ }\textbf {\bibinfo {volume}
  {79}},\ \bibinfo {pages} {235433} (\bibinfo {year} {2009})}\BibitemShut
  {NoStop}%
\bibitem [{\citenamefont {Zhang}\ and\ \citenamefont {Jo}(2015)}]{Zhang2015}%
  \BibitemOpen
  \bibfield  {author} {\bibinfo {author} {\bibfnamefont {T.}~\bibnamefont
  {Zhang}}\ and\ \bibinfo {author} {\bibfnamefont {G.-B.}\ \bibnamefont {Jo}},\
  }\href {\doibase doi:10.1038/srep16044} {\bibfield  {journal} {\bibinfo
  {journal} {Scientific Reports}\ }\textbf {\bibinfo {volume} {5}},\ \bibinfo
  {pages} {16044} (\bibinfo {year} {2015})}\BibitemShut {NoStop}%
\bibitem [{\citenamefont {Lifshitz}(1960)}]{Lifshitz60}%
  \BibitemOpen
  \bibfield  {author} {\bibinfo {author} {\bibfnamefont {I.~M.}\ \bibnamefont
  {Lifshitz}},\ }\href
  {http://www.jetp.ac.ru/cgi-bin/e/index/e/11/5/p1130?a=list} {\bibfield
  {journal} {\bibinfo  {journal} {Soviet Physics JEPT}\ }\textbf {\bibinfo
  {volume} {11}},\ \bibinfo {pages} {1130} (\bibinfo {year}
  {1960})}\BibitemShut {NoStop}%
\bibitem [{\citenamefont {Xia}\ \emph {et~al.}(2010)\citenamefont {Xia},
  \citenamefont {Farmer}, \citenamefont {Lin},\ and\ \citenamefont
  {Avouris}}]{Xia10}%
  \BibitemOpen
  \bibfield  {author} {\bibinfo {author} {\bibfnamefont {F.}~\bibnamefont
  {Xia}}, \bibinfo {author} {\bibfnamefont {D.~B.}\ \bibnamefont {Farmer}},
  \bibinfo {author} {\bibfnamefont {Y.-m.}\ \bibnamefont {Lin}}, \ and\
  \bibinfo {author} {\bibfnamefont {P.}~\bibnamefont {Avouris}},\ }\href
  {\doibase 10.1021/nl9039636} {\bibfield  {journal} {\bibinfo  {journal} {Nano
  Letters}\ }\textbf {\bibinfo {volume} {10}},\ \bibinfo {pages} {715}
  (\bibinfo {year} {2010})}\BibitemShut {NoStop}%
\bibitem [{\citenamefont {Yung}\ \emph {et~al.}(2013)\citenamefont {Yung},
  \citenamefont {Wu}, \citenamefont {Pierpoint},\ and\ \citenamefont
  {Kusmartsev}}]{Yung13}%
  \BibitemOpen
  \bibfield  {author} {\bibinfo {author} {\bibfnamefont {K.}~\bibnamefont
  {Yung}}, \bibinfo {author} {\bibfnamefont {W.}~\bibnamefont {Wu}}, \bibinfo
  {author} {\bibfnamefont {M.}~\bibnamefont {Pierpoint}}, \ and\ \bibinfo
  {author} {\bibfnamefont {F.}~\bibnamefont {Kusmartsev}},\ }\href {\doibase
  10.1080/00107514.2013.833701} {\bibfield  {journal} {\bibinfo  {journal}
  {Contemporary Physics}\ }\textbf {\bibinfo {volume} {54}},\ \bibinfo {pages}
  {233} (\bibinfo {year} {2013})}\BibitemShut {NoStop}%
\bibitem [{\citenamefont {Wakabayashi}\ \emph {et~al.}(2010)\citenamefont
  {Wakabayashi}, \citenamefont {ichi Sasaki}, \citenamefont {Nakanishi},\ and\
  \citenamefont {Enoki}}]{katsunori}%
  \BibitemOpen
  \bibfield  {author} {\bibinfo {author} {\bibfnamefont {K.}~\bibnamefont
  {Wakabayashi}}, \bibinfo {author} {\bibfnamefont {K.}~\bibnamefont {ichi
  Sasaki}}, \bibinfo {author} {\bibfnamefont {T.}~\bibnamefont {Nakanishi}}, \
  and\ \bibinfo {author} {\bibfnamefont {T.}~\bibnamefont {Enoki}},\ }\href
  {\doibase 10.1088/1468-6996/11/5/054504} {\bibfield  {journal} {\bibinfo
  {journal} {Science and Technology of Advanced Materials}\ }\textbf {\bibinfo
  {volume} {11}},\ \bibinfo {pages} {054504} (\bibinfo {year}
  {2010})}\BibitemShut {NoStop}%
\bibitem [{\citenamefont {Lieb}(1989)}]{Lieb89}%
  \BibitemOpen
  \bibfield  {author} {\bibinfo {author} {\bibfnamefont {E.~H.}\ \bibnamefont
  {Lieb}},\ }\href {\doibase 10.1103/PhysRevLett.62.1201} {\bibfield  {journal}
  {\bibinfo  {journal} {Phys. Rev. Lett.}\ }\textbf {\bibinfo {volume} {62}},\
  \bibinfo {pages} {1201} (\bibinfo {year} {1989})}\BibitemShut {NoStop}%
\bibitem [{\citenamefont {Zeller}(2006)}]{Zeller06}%
  \BibitemOpen
  \bibfield  {author} {\bibinfo {author} {\bibfnamefont {R.}~\bibnamefont
  {Zeller}},\ }\href
  {https://juser.fz-juelich.de/record/51139/files/NIC-Band-31.pdf} {\bibfield
  {journal} {\bibinfo  {journal} {NIC Series}\ }\textbf {\bibinfo {volume}
  {31}},\ \bibinfo {pages} {419} (\bibinfo {year} {2006})}\BibitemShut
  {NoStop}%
\bibitem [{\citenamefont {Koop}\ and\ \citenamefont {Wessel}(2017)}]{Wessel17}%
  \BibitemOpen
  \bibfield  {author} {\bibinfo {author} {\bibfnamefont {C.}~\bibnamefont
  {Koop}}\ and\ \bibinfo {author} {\bibfnamefont {S.}~\bibnamefont {Wessel}},\
  }\href {\doibase 10.1103/PhysRevB.96.165114} {\bibfield  {journal} {\bibinfo
  {journal} {Phys. Rev. B}\ }\textbf {\bibinfo {volume} {96}},\ \bibinfo
  {pages} {165114} (\bibinfo {year} {2017})}\BibitemShut {NoStop}%
\bibitem [{\citenamefont {Tomczak}\ \emph {et~al.}(2017)\citenamefont
  {Tomczak}, \citenamefont {Liu}, \citenamefont {Toschi}, \citenamefont
  {Kresse},\ and\ \citenamefont {Held}}]{Tomczak2017}%
  \BibitemOpen
  \bibfield  {author} {\bibinfo {author} {\bibfnamefont {J.~M.}\ \bibnamefont
  {Tomczak}}, \bibinfo {author} {\bibfnamefont {P.}~\bibnamefont {Liu}},
  \bibinfo {author} {\bibfnamefont {A.}~\bibnamefont {Toschi}}, \bibinfo
  {author} {\bibfnamefont {G.}~\bibnamefont {Kresse}}, \ and\ \bibinfo {author}
  {\bibfnamefont {K.}~\bibnamefont {Held}},\ }\href {\doibase
  10.1140/epjst/e2017-70053-1} {\bibfield  {journal} {\bibinfo  {journal} {The
  European Physical Journal Special Topics}\ }\textbf {\bibinfo {volume}
  {226}},\ \bibinfo {pages} {2565} (\bibinfo {year} {2017})}\BibitemShut
  {NoStop}%
\end{thebibliography}%

\end{document}